\documentclass[useAMS,usenatbib]{mnras}
 \usepackage{graphicx}
 \usepackage{graphics}
 \usepackage{multicol}
 \usepackage{subfig}
 \usepackage{etex}
 \usepackage{setspace}
 \usepackage{float}

\usepackage[american]{babel}
\usepackage[utf8]{inputenc}
\usepackage[T1]{fontenc}
\usepackage{ae,aecompl}

\title[Pulse phase dependence of emission lines
in 4U~1626--67]{Changes in the pulse phase dependence of X-ray emission lines in 4U 1626-67 with a torque reversal}
 \author[A. Beri et al.]
 {Aru~Beri,$^{1,2}$
  Biswajit~Paul,$^1$ Gulab~C.~Dewangan,$^3$ \\
 $^1$, Raman Research Institute, Sadashivnagar, C. V. Raman Avenue, Bangalore-560 080, India.\\
 $^2$, Royal~Society--SERB Newton International Fellow,~School of Physics and Astronomy,~\\
 University of Southampton,~Southampton,~Hampshire SO17~1BJ,~UK \\
 $^3$, Inter University Centre for Astronomy and Astrophysics, Post bag 4, Ganeshkhind, Pune, India.\\}

\begin{document}
\pagerange{\pageref{firstpage}--\pageref{lastpage}} 
\maketitle
\label{firstpage}
\begin{abstract}
We report results from an observation with the \emph{XMM-Newton}
observatory of a unique X-ray pulsar
4U~1626--67.~\textsc{EPIC}-pn data during the current spin-up phase of 4U~1626--67 have been used 
to study pulse phase dependence of low energy emission lines.
We found
strong variability of low energy emission line at 0.915~keV with the pulse phase,
varying by a factor of 2, much
stronger than the continuum variability.
Another interesting observation is that behavior of one of the low energy emission
lines across the pulse phase is quite different from that
observed during the spin-down phase.
This indicates that the structures in the accretion disk
that produce pulse phase dependence of emission features 
have changed from spin-down to spin-up phase.
This is well supported by the differences in the timing characteristics (like pulse profiles,~QPOs etc)  
between spin-down and spin-up phases.
 We have also found that during the current spin-up phase of 4U~1626--67, the X-ray pulse profile below 2~keV
is different compared to the spin-down phase.
The X-ray light curve also shows flares which produce a feature around 3~mHz
in power density spectrum
of 4U~1626--67.~Since flares are dominant
at lower energies, the feature around 3~mHz is prominent at low energies.

\end{abstract}
\section{Introduction}
4U~1626--67 is a remarkable ultra-compact X-ray binary bearing a neutron star with pulse period
of 7.7~seconds \citep{Rappaport77}.
Evidence of binary motion has never been revealed from X-ray timing measurements
\citep[see e.g.,][]{Rappaport77,Joss78,Jain08}.
 Orbital period of 42~minutes has been inferred from the 
pulsed optical emission reprocessed
on the surface of secondary
\citep{Middleditch81,Chakrabarty98}.
An upper limit of 10~lt-ms for pulse arrival delay 
has been reported by \citet{Jain08} using X-ray data from \emph{RXTE}-\textsc{PCA}.
Time-scales of torque reversals observed in most of the accretion powered pulsars
varies from weeks to months and years and, in most cases accretion torques
are often related to the X-ray luminosity. 
4U~1626--67, a persistent X-ray source 
underwent two torque reversals since its discovery \citep{Camero10}. 
It was initially observed in spin-up state, this trend
reversed in 1990 and the neutron star began to spin-down.
After the steady spin-down phase of about 18 years, a transition to spin-up 
took place in 2008.~The second torque reversal was detected with \emph{RXTE}-\textsc{PCA} \citep{Jain-atel09}
and \emph{Fermi}-\textsc{GBM} \citep{Arranz09}.
~Moreover, it is observed that this source does not obey standard X-ray luminosity-accretion torque
relation \citep{Beri14}.
X-ray features during spin-down phase were different in comparison 
with both spin-up phases.
The most outstanding difference in energy resolved pulse profiles
of the two spin-up eras and the spin down era was disappearance 
of the sharp double peaked profile during spin-down era \citep[for details see][]{Beri14}.
Quasi periodic oscillation~(QPO) at 48~mHz was observed in all the observations
during spin-down phase \citep{Kaur08},~this feature was absent in the power
density spectra (PDS) created using X-ray data during current spin-up phase \citep{Jain10}.

The X-ray spectrum of 4U~1626--67 is well described using two
continuum components: a hard power law and a black body.
X-ray spectra during both spin-up phases showed blackbody temperature of about 0.6~keV
while during the spin-down phase of 4U~1626--67 the blackbody temperature decreased to
$\sim$~0.3~keV.~Moreover, the energy spectrum became harder during the spin-down phase.
The power law photon index showed a value of $\sim$~1.5 during the first spin-up phase
which changed to $\sim$~0.4-0.6 during the spin-down phase and during the second spin-up phase
it showed a value in the range of 0.8-1.0
\citep[see][references therein]{Beri14}.
Detailed study of this source during each phase of torque reversal
suggests that accretion flow geometry is different during the spin-up and spin-down
phases and plays an 
important role in transfer of angular momentum \citep{Jain10,Beri14}. \\

X-ray spectrum of 4U~1626--67 is unique, unusually bright Neon~(Ne)
and Oxygen~(O) lines have been reported from many spectroscopic
observations \citep[][]{Angelini95,Owens97,Schulz01,Krauss07}.
Observations made with \emph{Chandra} revealed double peaked
nature of low energy emission line features, indicating their formation
in the accretion disk \citep{Schulz01}.
Continuum of the spectra is well described using a soft emission component and a
power law \citep[see][and references therein]{Beri15} though.

Observations made during spin-down and spin-up 
phase of 4U~1626--67 with the \emph{Suzaku} observatory were used to measure spectral changes
with torque reversal in 2008 \citep{Camero12}. The authors confirmed
that the equivalent width and the intensity of these emission 
lines are variable. They found that fluxes of all the emission lines
have increased almost by factor of $\sim$~5 with an exception
of Ne~X~(1.02~keV) emission line that showed an increase by factor of $\sim$~8
after the torque reversal.~Pulse
phase resolved spectroscopy performed using data from the \emph{XMM-Newton} observatory during 
spin-down phase of 4U~1626--67 revealed that line fluxes show pulse phase dependence \citep{Beri15}
One of the emission line (O~VII) showed the line flux to vary by a factor of about four,
significantly larger compared to
the relative variation of total flux.
Warp-like structures in the accretion disk are believed 
to be the cause of observed line flux variability. \\

An interesting possibility for the cause of spin-down is the radiation pressure induced warping of the inner
accretion disk which may become retrograde leading to negative accretion
torque \citep{Kerkwijk98}.
Moreover, changes in the timing characteristics~(like the pulse profile, the QPO's etc)
in the spin-down phase compared to spin-up phase
are understood to be due changes in the inner accretion flow
from a warped accretion disk in the spin-down phase.
Therefore, we expect to observe changes
in the accretion flow and probably 
also the accretion disk structures of 4U~1626--67 during
the spin-up phase.~We carried out a 
pulse phase resolved spectroscopy to investigate if this results
into a different modulation of the emission lines during its current spin-up phase.
In this paper we present results obtained from 
timing and spectral study of 4U~1626--67, performed using data 
obtained with the \emph{XMM-Newton} observatory during its current spin-up phase. 
The paper is structured as follows:~we describe observation
details and data reduction procedure in Section~2.~This is followed by 
the results from timing analysis (Section~3).~In Section~4 we present results from the spectral
analysis.~The last section~(Section~5) of the paper presents
results and discussions. \\
\section{Observations and Data Reduction}
We have obtained a 56~ks observation of 4U~1626--67 during its current spin-up phase with \emph{XMM-Newton}.
The observation was performed
on October~5,~2015 bearing an ID-0764860101. \\
\emph{XMM-Newton} satellite has three X-ray telescopes, each with an Europeon
photon imaging camera~(\textsc{EPIC}) at the focus \citep{Jansen01}.
Two of the \textsc{EPIC} imaging spectrometers use metal-oxide semiconductors~(MOS)
CCDs \citep{Turner01} and one used pn CCD \citep{Struder01}.~Reflection grating spectrometer~(RGS)
and optical monitor~(OM) are two other instruments on-board \emph{XMM-Newton} satellite.
\textsc{RGS} comprises of two spectrometers namely, \textsc{RGS1} and \textsc{RGS2}.
Two \textsc{RGS} have a bandpass of 0.35--2.5~keV 
and first-order spectral resolution of about 200 to 800 in 0.35--2.5~keV.~They are attached to two of the X-ray telescopes 
with MOS.~Simultaneous optical/UV observations
are carried out with the optical monitor~(OM). \\

In this work we performed analysis using data from 
the \textsc{EPIC}-pn and the \textsc{RGS} on-board \emph{XMM-Newton}.
\textsc{EPIC}-pn data were collected in timing mode using medium filter
with a frame time of 6~ms.~In timing mode only one CCD chip
is in operation and data is collapsed into one dimensional row
and read out at high speed.
~\textsc{RGS} data was operated in standard
$\textquoteleft$spectral' mode. \\
We processed the \emph{XMM-Newton} observation data files, using the
science analysis software~(SAS version 15.0).~Latest updated calibration
files available as on April 2016 were applied. \\
~Standard SAS
tool \textsc{epproc} was used to obtain \textsc{EPIC}-pn event file.
We first checked for flaring particle background in the data.
A light curve was extracted using selection criterion:~PATTERN=0
in the energy range of 10-12~keV.
We found no evidence of soft proton flaring.~Thereafter, we extracted
\textsc{EPIC}-pn cleaned event list by selecting events with PATTERN$\leq$4,
FLAG=0 and energy in the range 0.3-12~keV.~This cleaned event file was used to 
extract source events and background files.~We used rectangular box with RAWX~=~
30-46 for source events and RAWX~=~2-4 for background.~Source event file was also checked
for photon pile-up using SAS tool \textsc{epatplot}.~No significant photon
pile-up was found.~Barycenter correction was performed using SAS tool \textsc{barycen}.
~For extraction of light curves and spectra SAS tool \textsc{evselect}
was used.~Response matrix and Ancillary response files were generated
using the SAS task \textsc{rmfgen} and \textsc{arfgen}, respectively. \\
For the \textsc{RGS} data reduction, we used SAS tool \textsc{rgsproc}
to reduce and extract calibrated source and background spectrum and response files.
Standard procedure as mentioned in SAS analysis thread was followed. \\

\section{Timing Analysis}
\begin{figure*}
\centering
\begin{minipage}{0.45\textwidth}
\includegraphics[height=4.0in,width=6.0in,angle=0, keepaspectratio]{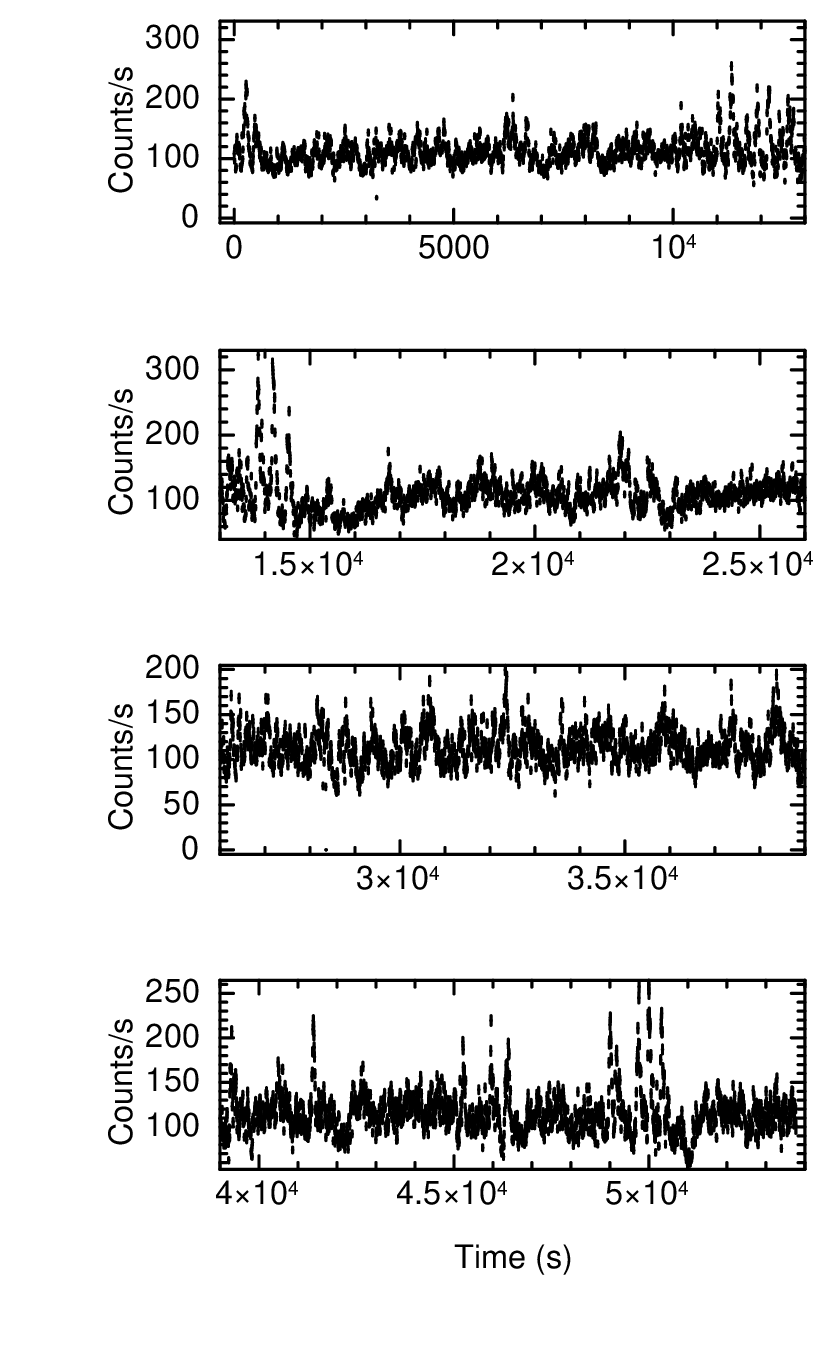}
\end{minipage}
\hspace{0.05\linewidth}
\begin{minipage}{0.45\textwidth}
 \includegraphics[height=3.5in,width=8.5cm]{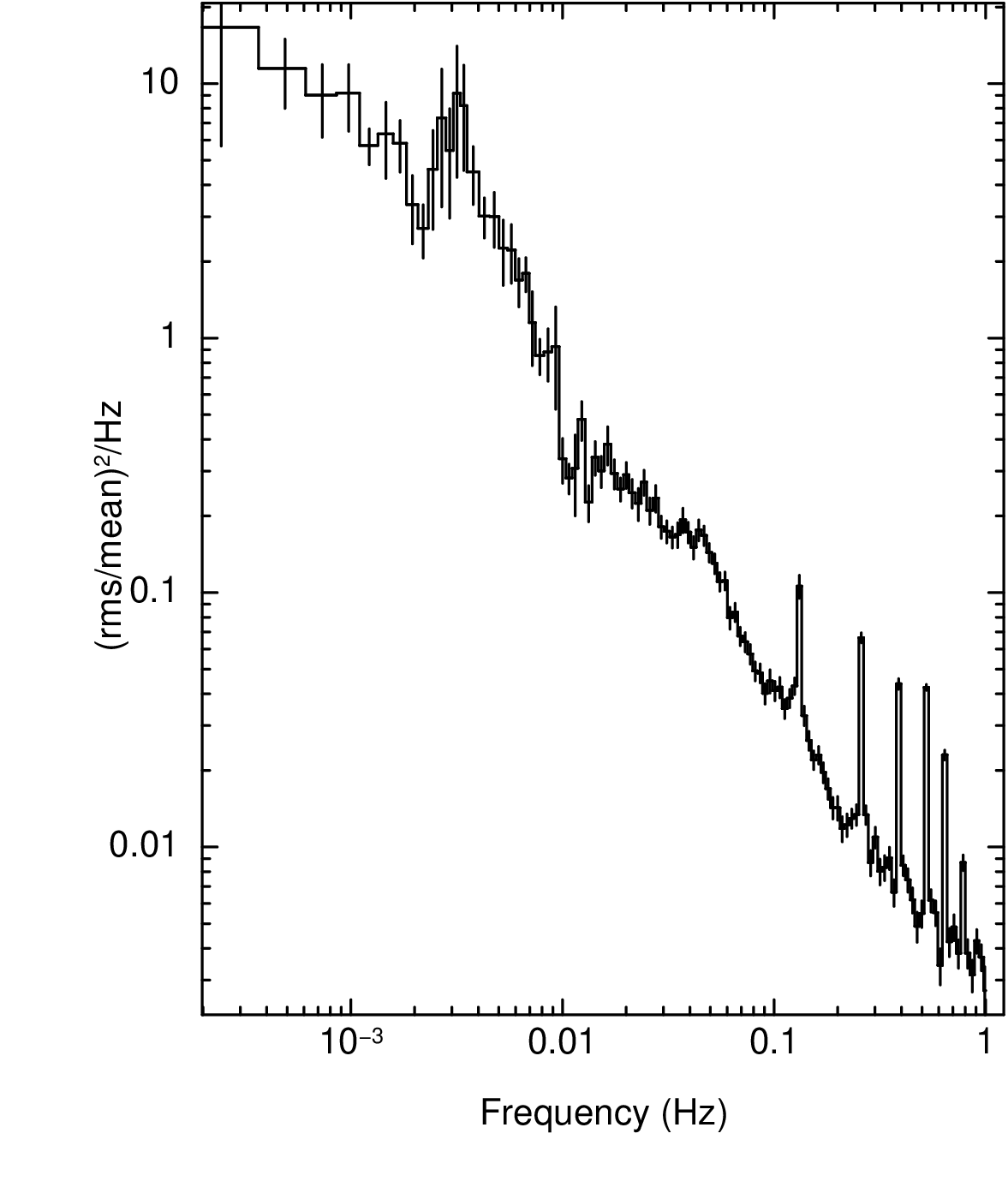}
\end{minipage}
\caption{Light curves in 0.3-12~keV band of 4U~1626--67, created using data from \textsc{EPIC}-pn on-board \emph{XMM-Newton} are shown in left. 
We divided data into four short segments each of $\sim$~14~ks for visual clarity of flares.~Light curves are binned using 7.7~seconds.~In the right, we show power density spectrum~(PDS) created using the same light curve.}
\label{LC}
\end{figure*}

Left hand side of Figure~\ref{LC} shows barycenter corrected and background subtracted 
light curve of 4U~1626--67 obtained using the \textsc{EPIC}-pn data.
Light curve is highly variable.
It includes both flares and dip features in it.
X-ray flares have also been observed in the light curves created using previous observations
made during its spin-up phase \citep[see][and references therein]{Beri14}. 
Amplitude of flares is 2-3 times above the persistent level (Figure~\ref{LC}).
Duration of flares is few hundred of seconds.
Recurrence time-scales of these flares varies between 300 to 1000~seconds
and these time-scales are consistent with the previous reports \citep[see][]{Joss78,Li80,Raman16} \\
~Unlike other flaring sources
 like LMC~X-4, SMC~X--1,~where persistent emission begins just after the end of flares,
it is interesting to notice a sharp dip in the light curve near the decay of bright flares
 at 18000,~23000~seconds~(second panel) and 51000~seconds~(fourth panel)
 of Figure~\ref{LC}.
This feature has never been reported before in 4U~1626--67.
~Similar kind of dip near the end of outburst has also
been observed in the light curves of bursting pulsar~(GRO~J1744-28)
\citep[eg.,][]{Giles96}. 
\subsection{Power density Spectrum}
\begin{figure*}
\centering
\begin{minipage}{0.45\textwidth}
\includegraphics[height=3.75in,width=7.5cm]{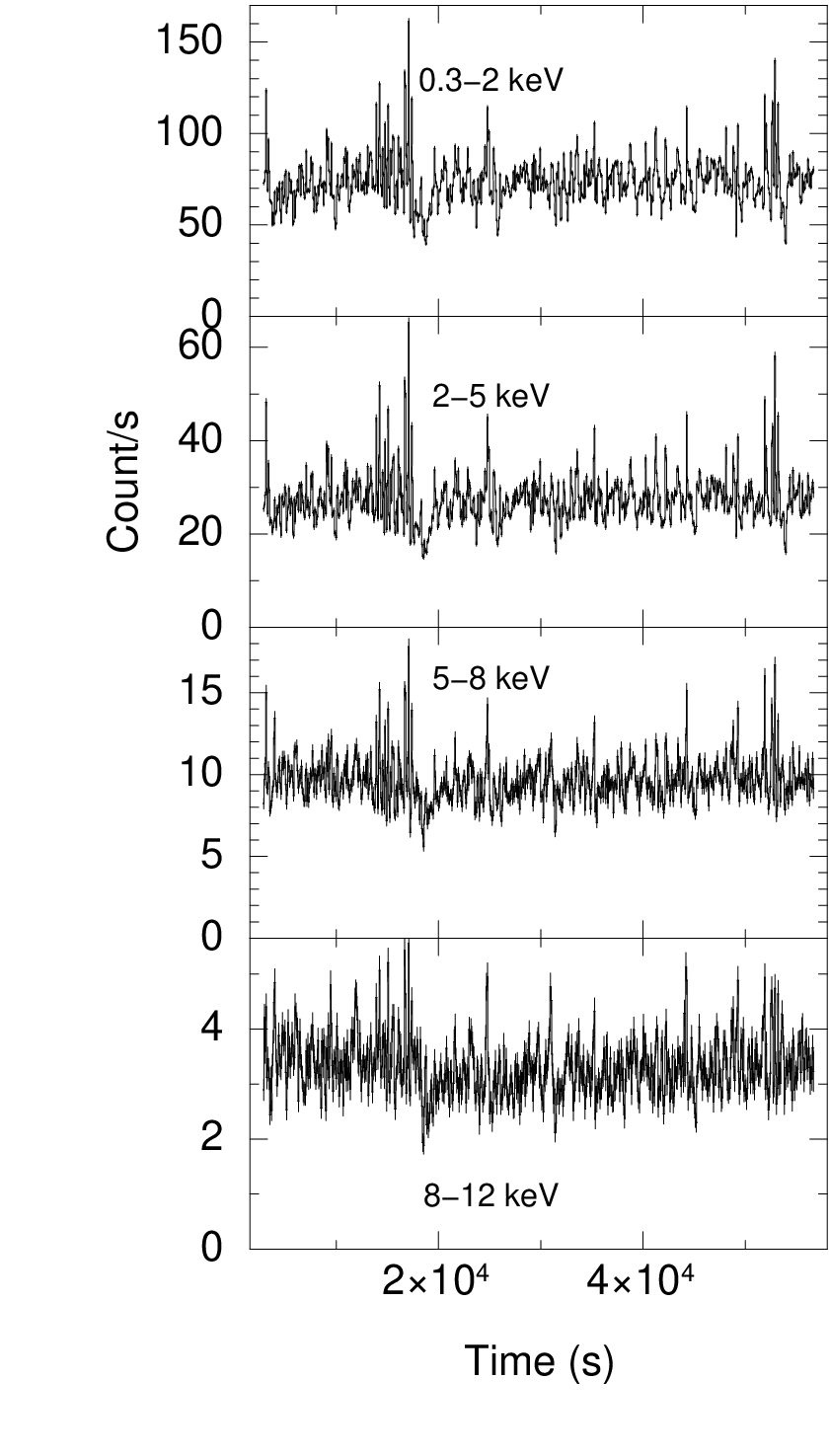}
\end{minipage}
\hspace{0.05\linewidth}
\begin{minipage}{0.45\textwidth}
\includegraphics[height=3.75in,width=7.5cm]{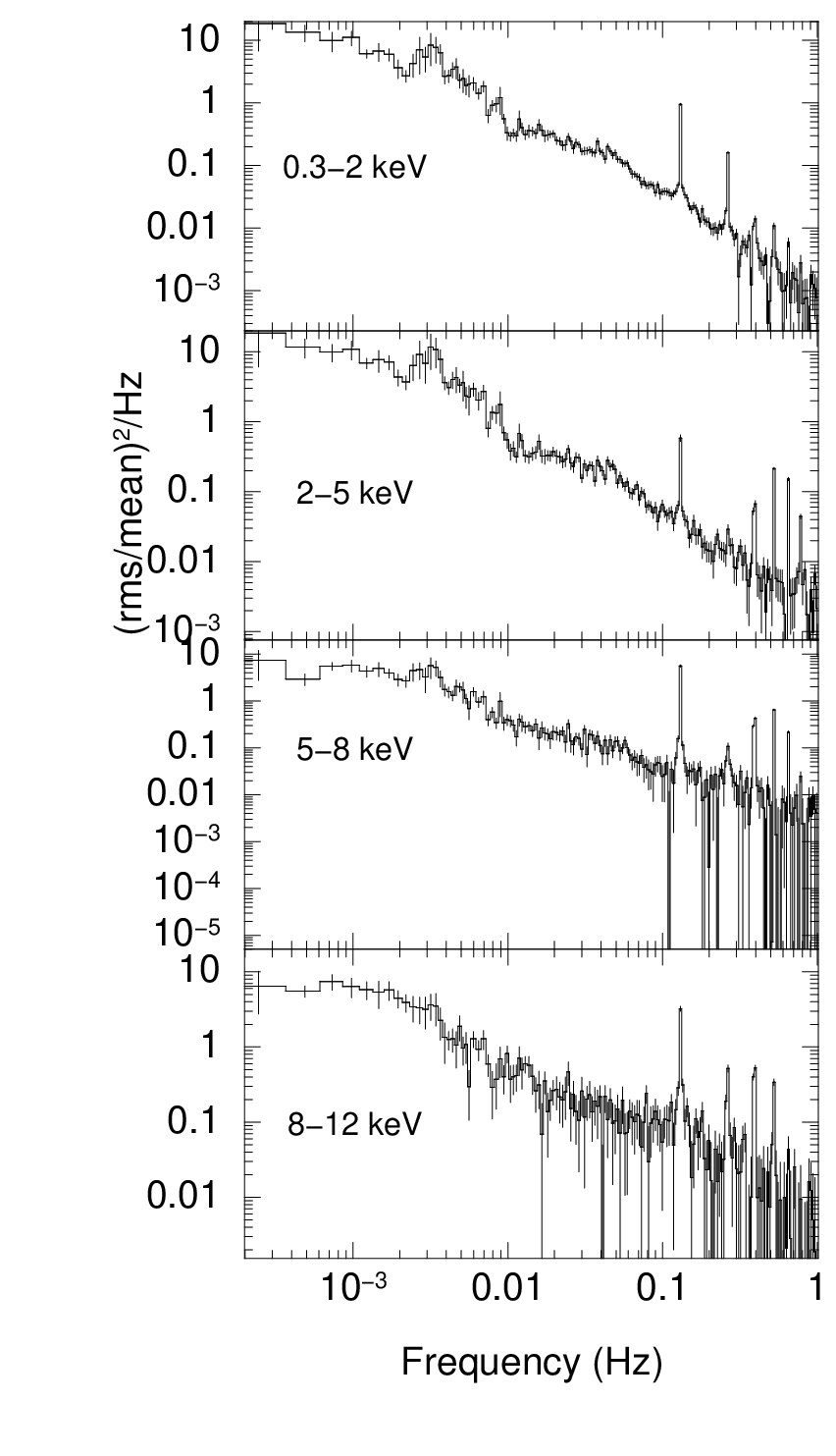}
\end{minipage}
\caption{Energy resolved light curves binned with 77~seconds are shown in left.~These light curves 
were used to generate power density spectra, shown in the right.
This plot shows a QPO feature at $\sim$~3~mHz and some energy dependence of Power Density Spectrum of 4U~1626--67.}
\label{PDS_EN}
\end{figure*}

Power density spectrum~(PDS) generated using the \textsc{EPIC}-pn light curve
is shown in Figure-\ref{LC}. The light curve was divided into stretches of 8192 seconds. PDS from all the segments were
averaged to produce the final PDS and were normalized such that their integral
gives squared rms fractional variability and the white noise level was
subtracted. PDS showed narrow peak at around 0.130 Hz which corresponds to the
spin frequency of the neutron star. Multiple harmonics are also seen
in the PDS of the source.~In addition to the main peak,~a QPO
feature is seen at $\sim$~3~mHz with fractional rms amplitude of $\sim$~$7.26\pm0.07~{\%}$.
3~mHz QPO can be due to flares seen in the light curve
and this feature is observed for the first time in the X-ray data during current 
(spin-up) phase of 4U~1626--67.~A similar mHz QPO was however observed in the PDS generated with X-rays
during the first spin-up phase of 4U~1626--67 \citep[see e.g.,][]{Joss78}.
Another interesting observation is dependence of 3~mHz QPO on energy~(Figure~\ref{PDS_EN}).
It is evident from Figure~\ref{PDS_EN} that flares are more prominent 
at lower energies and therefore sharp feature
in the PDS around 3~mHz is dominant at energies below 5~keV.~Fractional rms amplitude of 3~mHz QPO feature in the PDS created using 
light curves in different energy bands namely, 0.3-2~keV,~2-5~keV,
5-8~keV amd 8-12~keV are $7.8\pm0.1~{\%}$,~$9.4\pm0.4~{\%}$,
$4\pm2~{\%}$,~$4\pm2~{\%}$ respectively.~We also detect a signature of broad QPO around 48~mHz with rms amplitude 
of $\sim$~5~{\%}, much smaller than the rms in the 48 mHz QPO seen during the spin-down phase \citep[e.g.,][]{Kommers98}. 

\begin{figure*}
\centering
\begin{minipage}{0.45\textwidth}
\includegraphics[height=3.5in,width=7.5cm]{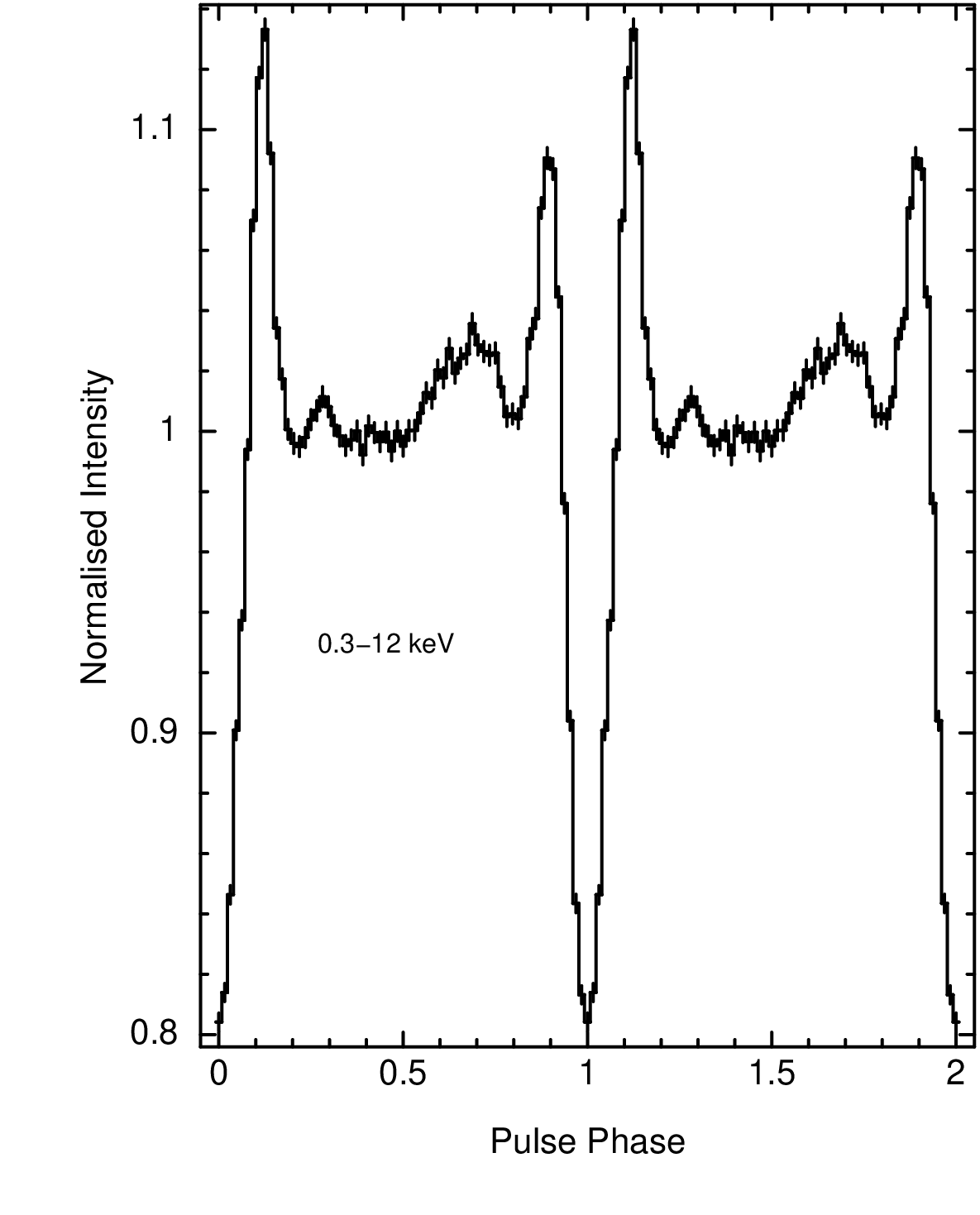}
\end{minipage}
\hspace{0.05\linewidth}
\begin{minipage}{0.45\textwidth}
\includegraphics[height=3.5in,width=7.5cm]{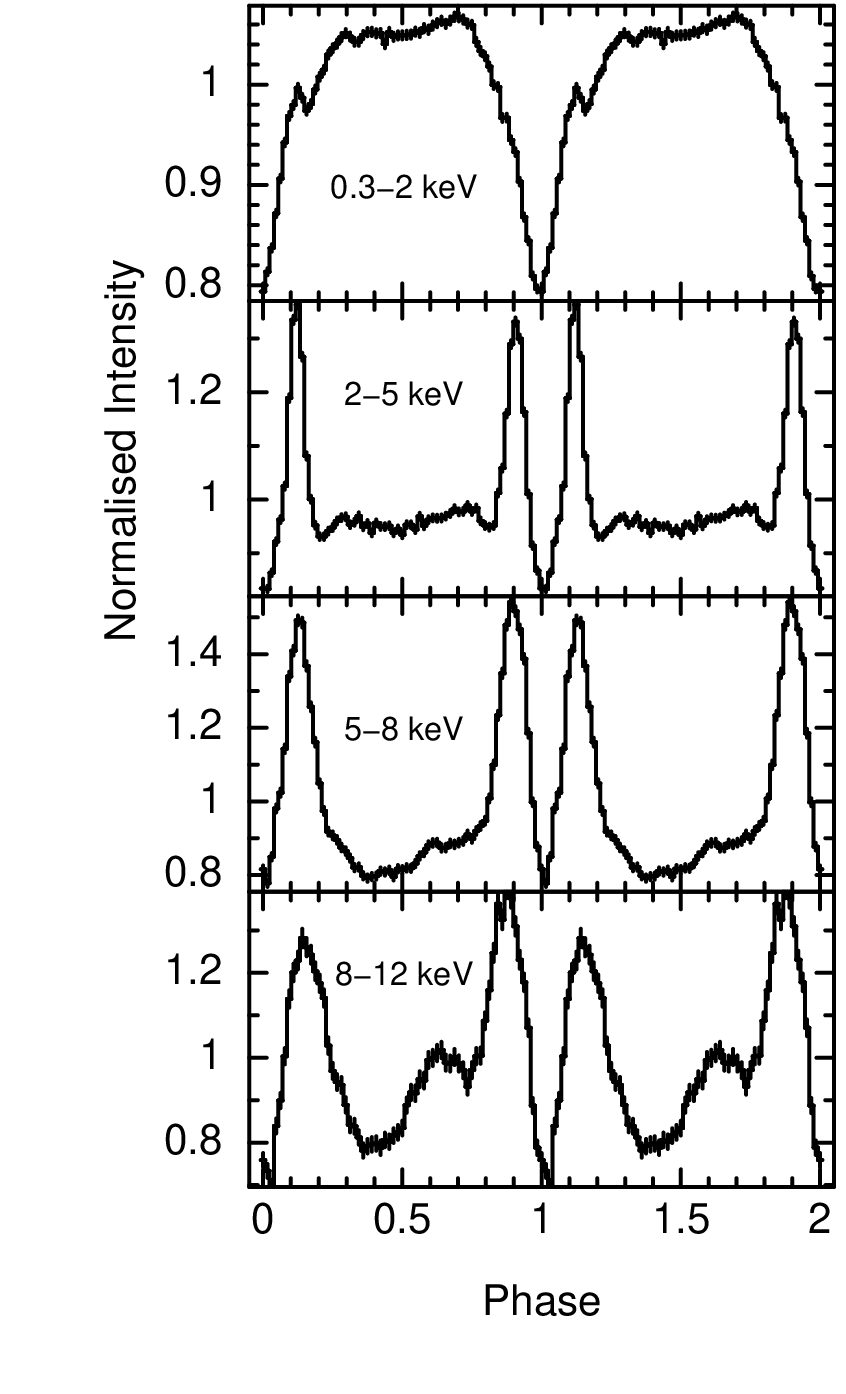}
\end{minipage}
\caption{Left:~The average pulse profile created in the energy band 0.3-12~keV.~Right:~The
energy resolved pulse profiles.~Pulse profiles are binned into 64 phasebins.}
\label{EN-pp}
\end{figure*}

\begin{figure*}
\centering
\includegraphics[height=3.5in,width=9.5cm]{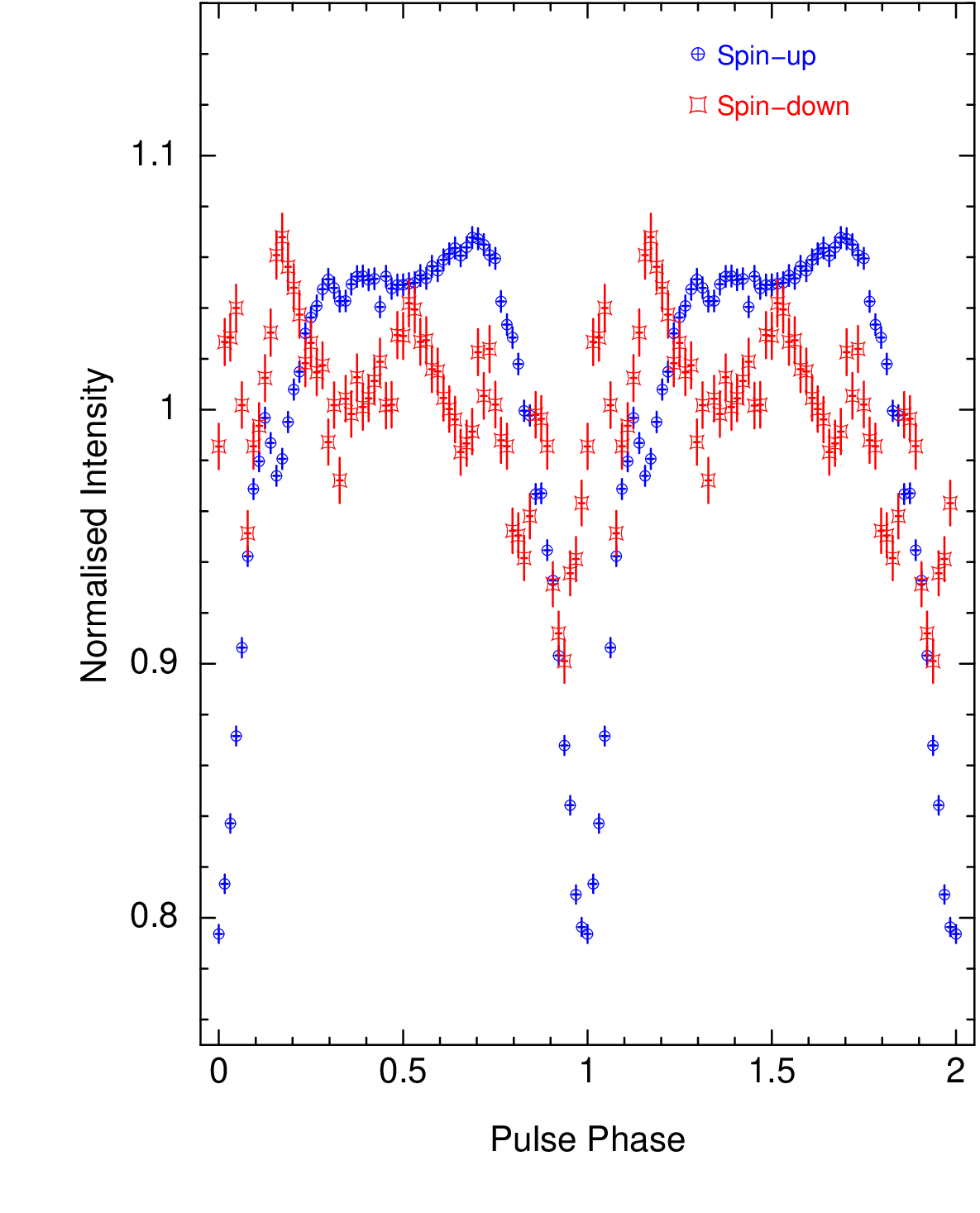}
\caption{Pulse Profile created in the energy band of 0.3-2~keV using \emph{XMM-Newton}-pn data of spin-down is shown in red.
~This figure demonstrates that pulse profile below 2~keV is quite different from that seen in the current spin-up phase~(blue).}
\label{PP-XMM}
\end{figure*}

\subsection{Pulse Profiles}
Spin-period was determined to be $7.67255\pm0.00009$~seconds using epoch folding $\chi^2$ 
maximization technique.
This period was used for creating pulse profiles.
We first created average pulse profile with 64 phase~bins,~using light curve in 
0.3-12~keV energy band~(Figure~\ref{EN-pp}).
The bi-horned peaks observed in the pulse profiles 
look similar to the previously reported pulse profiles
of 4U~1626--67 during its spin-up era \citep[see e.g.,][]{Beri14}.
However,~in 0.3-12~keV band amplitude of first peak is slightly small compared to the second peak. \\

The energy resolved pulse profiles were created using the light curves in 
the energy bands of 0.3-2~keV,~2-5~keV,~5-8~keV,~8-12~keV.
Thanks to \emph{XMM-Newton} which enabled us to investigate pulse profiles below 2~keV
during its current spin-up phase for the first time.
Pulse profile in 0.3-2~keV band looks simple having shoulder-like
structure.~It has a sharp dip around phase 0.0.
It seems that the sharp dip observed in the 0.3-2~keV profiles mainly contributes
to the dip observed between the two horns in the energy
averaged profiles~(0.3-12~keV).
Profile shape in 0.3-2~keV band is similar to that seen 
during the first spin-up phase \citep{Pravdo79}. 
It is interesting
to see that pulse profile below 2~keV is quite different from that observed
in other energy bands.~This suggests that pulsation of soft component is different from
that in higher energy band pulse profiles
which indicate that pulsation of thermal component
is different from power law component.~A soft spectral component, that pulsates differently
from the power law component has been detected in other sources with low absorption column density
~(e.g., SMC~X--1 and LMC~X--4) and have been interpreted as reprocessed thermal emission from 
the inner accretion disk \citep{Paul02}.

To compare pulse profiles below 2~keV during its current phase
with that of the spin-down phase, we created pulse profile in the energy band
of 0.3-2~keV using data from previous \emph{XMM-Newton} observation~(ObsID-0152620101)
during its spin-down phase.~We performed data reduction and analysis
in the same way as discussed in our previous paper on 4U~1626--67 
\citep{Beri15}.
It is interesting to see that pulse profile during spin-down phase
is quite different from that during its current phase.
Pulse profile below 2~keV has many structures during spin-down phase
which is not the case during its current spin-up phase (see Figure~\ref{PP-XMM}).
Pulse profiles in remaining energy bands are consistent
with previous observations in spin-up state \citep[see][and references therein]{Beri14}. \\

\begin{figure*}
\centering
\includegraphics[height=4.in,width=9.cm,angle=270,keepaspectratio]{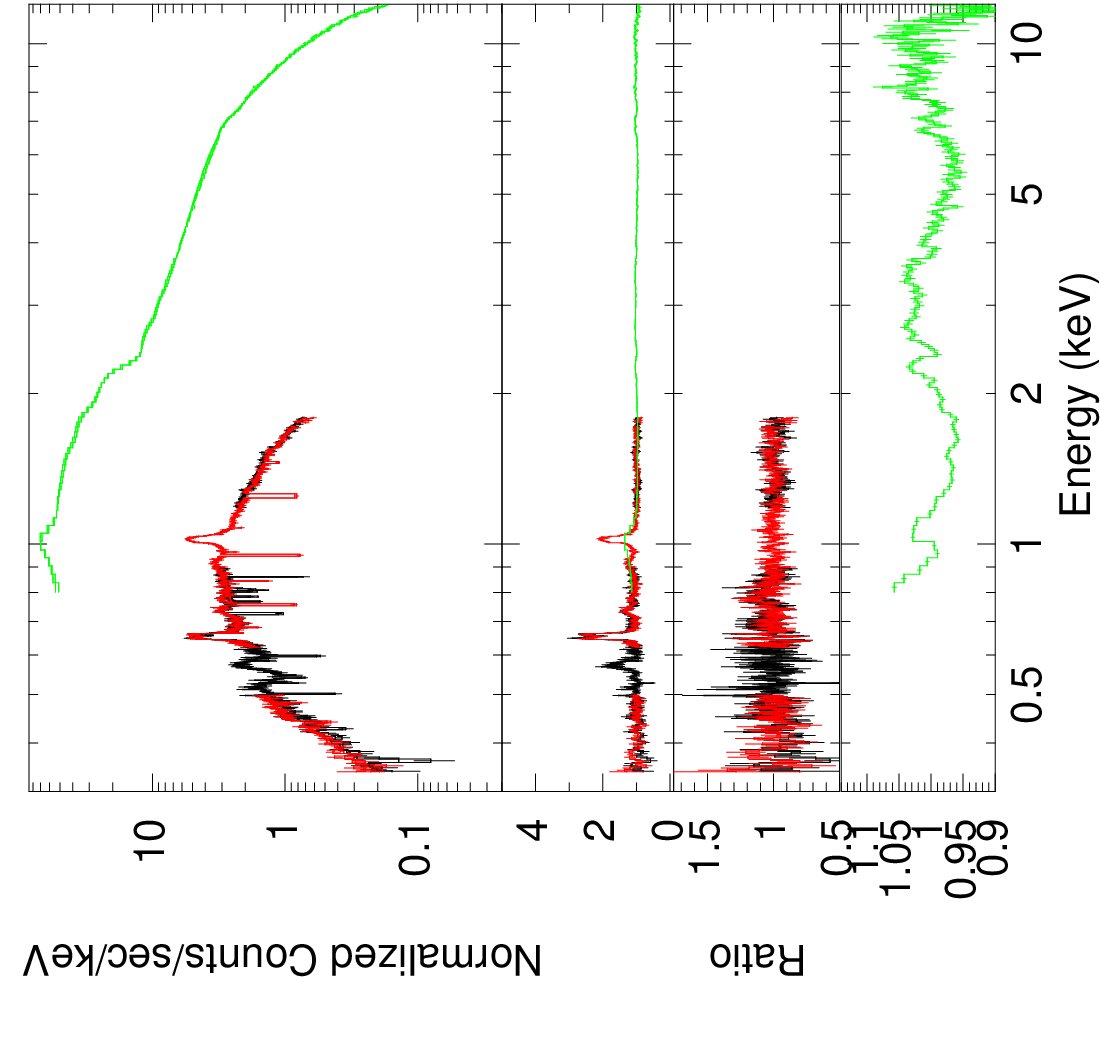}
\caption{Best fitted phase averaged spectrum obtained after performing simultaneous fit of \textsc{RGS} and \textsc{EPIC}-pn data.~Second
panel shows the ratio plot obtained after adding only continuum components while bottom two panels
show the ratio plots of pn and \textsc{RGS} respectively, obtained after the best fit.}
\label{Avg-spec-XMM}
\end{figure*}

\begin{table*}
\centering
      	 \caption{Best-fitting parameters obtained from simultaneous fit of \textsc{RGS1}, \textsc{RGS2} \& pn Spectrum.}
         \begin{tabular}{ l  l }
         \hline
         \hline
           
         Parameter & Model Values  \\ 
         N$_H$ (10${^2}{^2}$atoms cm$^{-2}$) & $0.085\pm{0.003}$  \\
         $kT_{bbodyrad}~(keV)$    & $0.427\pm{0.005}$ \\
         PowIndex ($\Gamma$)  & $0.914\pm{0.014}$ \\ 
         $N_{PL}^a$  & $0.0229\pm{0.0006}$  \\
         $\rm{Line Energy^b}$  \\
         $\rm{O~VII}$ & $0.571\pm{0.001}$ \\
         $\rm{O~VIII}$& $0.6536\pm{0.0003}$ \\
         $\rm{Ne~IX}$ &  $0.913\pm{0.004}$ \\
         $\rm{Ne~X}$   & $1.022\pm{0.001}$ \\
         $\rm{Fe~(L shell)}$  & $0.733\pm{0.001}$ \\
         $\rm{Fe~(K shell)}$  & $6.8\pm{0.1}$ \\
         $\rm{Line Width^c}$  \\
         $\rm{O~VII}$ & $0.009\pm{0.001}$ \\
         $\rm{O~VIII}$& $0.0075\pm{0.0002}$ \\
         $\rm{Ne~IX}$ &  $0.035\pm{0.005}$ \\
         $\rm{Ne~X}$   & $0.0115\pm{0.0006}$ \\
         $\rm{Fe~(L shell)}$ & $0.009\pm{0.002}$ \\
         $\rm{Fe~(K shell)}$  & $0.14\pm{0.09}$ \\
         
         $\rm{Line Flux^d}$    \\        
         $\rm{O~VII}$ & $11\pm{1.}$ \\
         $\rm{O~VIII}$& $25.\pm{1.}$ \\
         $\rm{Ne~IX}$ &  $13\pm{1}$ \\
         $\rm{Ne~X}$   & $23.0\pm{1.}$ \\
         $\rm{Fe~(L shell)}$  & $7.\pm{0.7}$ \\
         $\rm{Fe~(K shell)}$  & $0.6\pm{0.3}$ \\
        
          
        Reduced $\chi^2$  &   1.43(dof 1337)   \\
  \hline
  \end{tabular}

 \bigskip

{{\bf{ Note}}: Errors quoted are with 90 $\%$ confidence range. \\
    \hspace{2.5in}         Energy range used is 0.35-1.8~keV for \textsc{RGS1} and \textsc{RGS2} and 0.8-12.0~keV for \textsc{EPIC}-pn. \\ 
   \hspace{2.6in}   a $\rightarrow$ Powerlaw normalisation~($N_{PL}$) 
       is in units of $\rm{photons~cm^{-2}~s^{-1}~keV^{-1}}$ at 1~keV \\
             b $\rightarrow$ Line Energy  in units of keV.  \\
             c $\rightarrow$ Line width in units of keV.  \\
          \hspace{1.65	in}  d $\rightarrow$ Gaussian normalisation is in units of $10^{-4}$$\rm{photons~cm^{-2}~s^{-1}}$ }\\
            
           \label{Best-fit}  
            \end{table*}

  \begin{table*}
\caption{DOUBLE-GAUSSIAN EMISSION LINE FITS}
\label{table2}
\begin{tabular}{ c c c c c c c c c}
\hline
\hline
Observatory & MJD 
& \multicolumn{2}{c} {Blueshifted Lines} 

& \multicolumn{2}{c} {Redshifted Lines}  \\

 & & V~($km~s^{-1})$ & $Flux^{a}$  
& V~($km~s^{-1})$ & $Flux^{a}$ & Reference  \\
\hline
& & &  O~VIII~(0.653~keV) & & & \\
\hline
\emph{Chandra} &  51803.6 & 1740$\pm$440 & 14.04$\pm$2.52 & 1900$\pm$480 &
17.82 $\pm$ 0.57   & \citet{Schulz01}   \\
\emph{XMM-Newton} &  52145.1 & 1930$\pm$260 & $17.6_{-4.4}^{4.7}$ & 1930$\pm$260 &
$21.9_{-4.5}^{+4.9}$ & \citet{Krauss07}  \\
\emph{Chandra} & 52795.1 & 1770$\pm$330 & $13.0_{-4.9}^{+5.7}$ & 1770$\pm$330 &
 $13.7_{-5.0}^{+5.8}$ &  \citet{Krauss07} \\
\emph{XMM-Newton} & 52871.2 & 1810$\pm$180 & $12.7_{-1.9}^{+2.1}$ & 1810$\pm$180 &
$12.4_{-1.8}^{+2.0}$ &  \citet{Krauss07} \\ 
\emph{XMM-Newton} & 56397 & 1535$\pm$158 & 71$\pm$6 & 1306$\pm$158 &
61 $\pm$ 6 & Current Work \\
\hline
& & & 	 Ne~X~(1.02~keV) & & &  \\
\hline
\emph{Chandra} & 51803.6  & 2220$\pm$350 & 8.15$\pm$0.93 & 1240$\pm$220  &
15.04 $\pm$ 1.65  & \citet{Schulz01} \\
\emph{XMM-Newton} & 52145.1  & 1910$\pm$450 & $11.1_{-4.6}^{+4.8}$ & 1910$\pm$450 &
$14.4_{-5.2}^{-3.4}$ & \citet{Krauss07} \\
\emph{Chandra} & 52795.1  & 1670$\pm$180 & $8.2_{-1.5}^{+1.4}$ &  1670$\pm$180 &
$10.5_{-1.6}^{+1.8}$ &  \citet{Krauss07} \\
\emph{XMM-Newton} & 52871.2 & 1780$\pm$420 & $11.3_{-2.6}^{2.6}$ & 1780$\pm$420 &
$9.0_{-2.6}^{2.5}$ &  \citet{Krauss07} \\
\emph{XMM-Newton} & 56397 & 1731$\pm$247 & 65$\pm$20 & 1484$\pm$247 &
98 $\pm$ 22  &  Current Work \\
\hline\\
\end{tabular}
\label{Velocity}
\\{Note:~a~$\rightarrow$ The Gaussian normalization is in units of $10^{-5} photons~cm^{-2}~s^{-1}$ }
\end{table*}

\section{Spectroscopy}

\subsection{Phase Averaged Spectroscopy}
We performed simultaneous spectral fitting, using data from \textsc{RGS} and \textsc{EPIC}-pn (Figure~\ref{Avg-spec-XMM}).
Spectra of 1st order obtained using \textsc{RGS1} and \textsc{RGS2}
were grouped using the tool \textsc{grppha} (\textsc{HEASOFT} \textit{Version}-6.17)
to contain 6 channels per bin.~We have used 0.35-1.8~keV band of \textsc{RGS} for spectral fitting.
~Mean spectrum extracted using \textsc{EPIC}-pn was
rebinned using the SAS task \textsc{specgroup} to oversample the \textsc{FWHM} of energy resolution by factor of 3 
and to obtain minimum of 25~counts/bin.
There is no reliable calibration below 0.7~keV for \textsc{EPIC}-pn
in timing mode \footnote{http://xmm2.esac.esa.int/docs/documents/CAL-TN-0018.pdf}
and the disagreement between the \textsc{EPIC}-pn and the \textsc{RGS}
is larger below 0.7~keV.~Therefore,
we have used 0.8-12~keV band of \textsc{EPIC}-pn for spectral fitting.
All the spectral parameters other than the relative instrument normalization,
were tied together for both \textsc{RGS} and \textsc{EPIC}-pn.
We fixed the instrumental normalization of \textsc{RGS1}
to 1, and freed the normalization of the \textsc{RGS2} and \textsc{EPIC}-pn
instruments.~The values of \textsc{constant} model component
obtained for \textsc{RGS2} and \textsc{EPIC}-pn are $0.977\pm0.007$
and $1.04\pm0.01$ respectively.
A blackbody component and a power law well describes 
the continuum of the phase-averaged spectrum \citep{Pravdo79,Kii86,Angelini95,Owens97,Orlandini98,Schulz01,Krauss07,Jain10,Iwakiri12}.
Therefore,~we modelled the continuum of the spectrum using \textit{tbabs*(bbodyrad+powerlaw)}. 
 Using only the continuum model showed a
significant excess in the residuals in the form of emission lines.
The second panel of Figure~\ref{Avg-spec-XMM}
shows the ratio between data and model, indicating the presence of low energy emission lines~(below~1~keV).
The raw \textsc{RGS} spectrum shows the presence of two strong emission lines around
0.65~keV and 1.0~keV, therefore, we added two Gaussian components around
 0.65 and 1.0~keV.~These line energies correspond to Ne~X and O~VIII.
 Adding these two Gaussian components was not adequate to obtain a good spectral fit.
 The presence of additional emission features at 0.73~keV, 0.571~keV, and 0.913~keV
 was observed in the 
residuals of the \textsc{RGS} data.
The presence of O~VII and Ne~IX emission lines around 0.569~keV and 0.915~keV respectively
in the X-ray spectrum of 4U~1626-67 have been reported earlier by several authors
\citep[see e.g.,][]{Schulz01,Krauss07}.
Therefore, to obtain an appropriate fit 
we added two additional Gaussian components at these line energies.
However, we required an additional Gaussian component
to model the excess seen around 0.73~keV.
This line energy correspond to iron~(Fe-L shell) emission feature.
Table~\ref{Best-fit} shows the best fit parameters obtained.~They are consistent with the
previous results during spin-up phase of 4U~1626--67 \citep[see e.g.,][]{Camero10}.
The equivalent widths~(EW) of O~VII~(0.571~keV),~O~VIII~(0.653~keV),~Fe-L shell~(0.73~keV),~Ne~IX~(0.913~keV) and 
Ne~X~(1.02~keV) emission lines are
$17.0\pm1.0$~eV,~$34\pm2$~eV,~$9.0\pm1.0$~eV,~$20\pm3.0$~eV,~$42.0\pm$1~eV respectively.\\
Here, we emphasize that for the first time the X-ray spectrum of
4U~1626-67 showed an Fe-L shell fluorescence emission
feature and the detection of this feature at 0.73~keV is statistically significant
as the value of chi-squared~($\chi^2$) increased from 2049 to 2458 (1341 degrees of freedom)
on fixing its normalization to zero.
 A systematic error of $2\%$ was added quadratically to each energy bin
to account for all the artifacts due to calibration issues in the \textsc{EPIC}-pn timing mode data.
The residuals of \textsc{EPIC}-pn showed the presence of a weak iron fluorescence emission line around 6.8~keV.
Therefore, we added another Gaussian component 
with line energy centered around 6.8~keV to the spectrum.~Equivalent
width of the emission line observed at 6.8~keV is $\sim$~$0.02\pm0.01$~keV.
The spectral fit resulted into reduced $\chi^2$~($\chi^2_{\nu}$) of 1.43 for 1337 degrees
of freedom (see Table-\ref{Best-fit}).
We also observed that on fixing the normalization of Fe K shell line
to zero lead to an increase in the value of chi-squared~($\chi^2$)
from 1914 to 1930 (1338 degrees of freedom) which suggests that the detection of this emission feature is statistically significant.
The presence of Fe~$K_{\alpha}$ emission line was also observed in the MOS~2 spectrum.
The values of line flux and the equivalent width observed in the MOS~2 data
are $0.58^{0.5}_{-0.3}$$\times$$10^{-4}$$\rm{photons~cm^{-2}~s^{-1}}$ and $0.015\pm0.010$~keV respectively.
These values are similar to that observed in 
the pn data and on fixing the normalization to zero of this line in the MOS~2 data also showed increase in 
the value of chi-squared~(172 to 183 for 171 degrees of freedom) which is similar to that observed in the pn data.
The addition of the systematic error to the pn data
is not likely to introduce any pulse phase dependence of emission line fluxes which is the main
motivation of the present work.~Here, we would like to mention that 
owing to limited statistical significance of Fe~$K_{\alpha}$ emission line we have not performed 
phase resolved spectroscopy for this line. \\

\begin{figure*}
\centering
\begin{minipage}{0.45\textwidth}
\includegraphics[height=4.in,width=6.5cm,keepaspectratio]{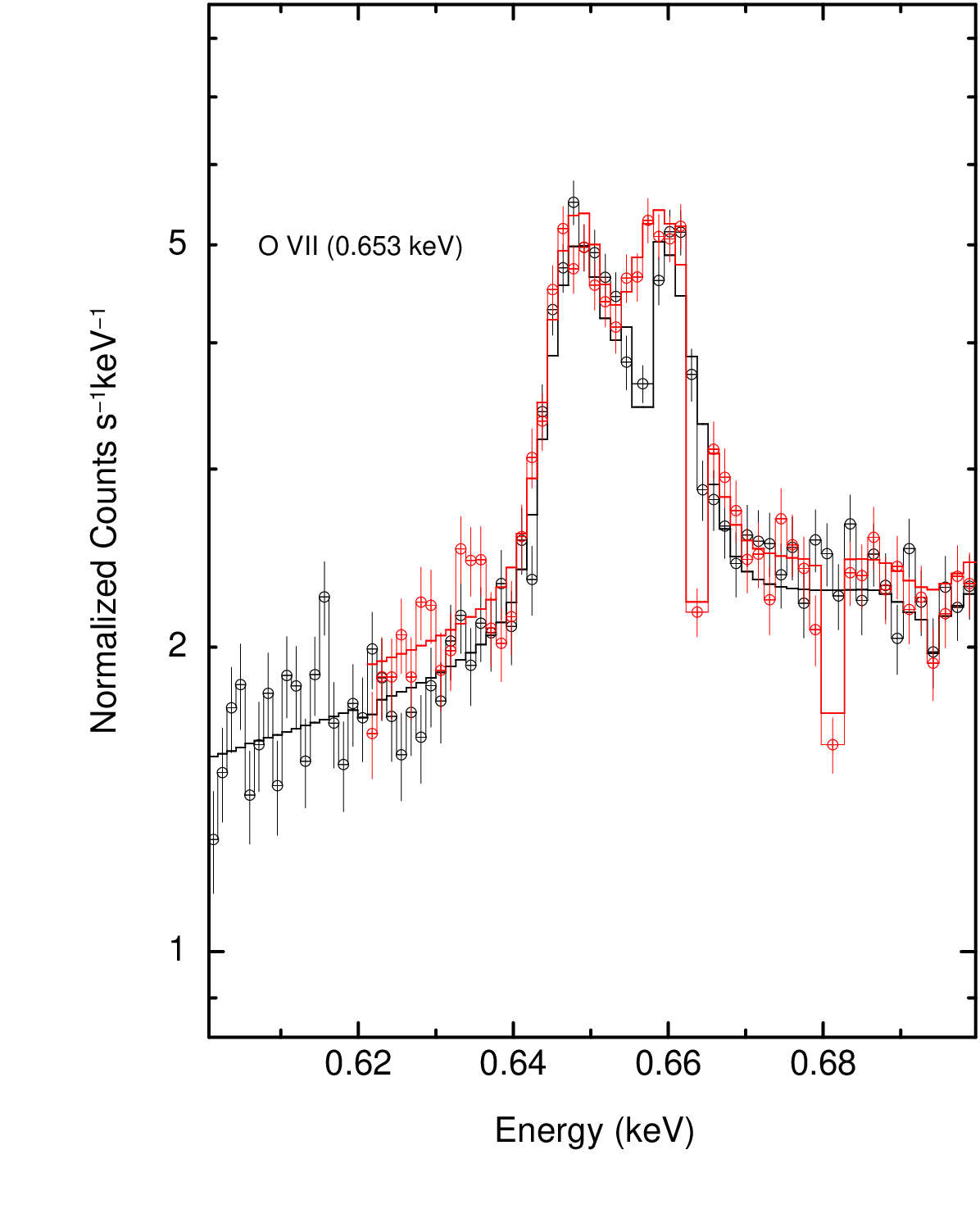}
\end{minipage}
\hspace{0.05\linewidth}
\begin{minipage}{0.45\textwidth}
\includegraphics[height=4.in,width=6.5cm,keepaspectratio]{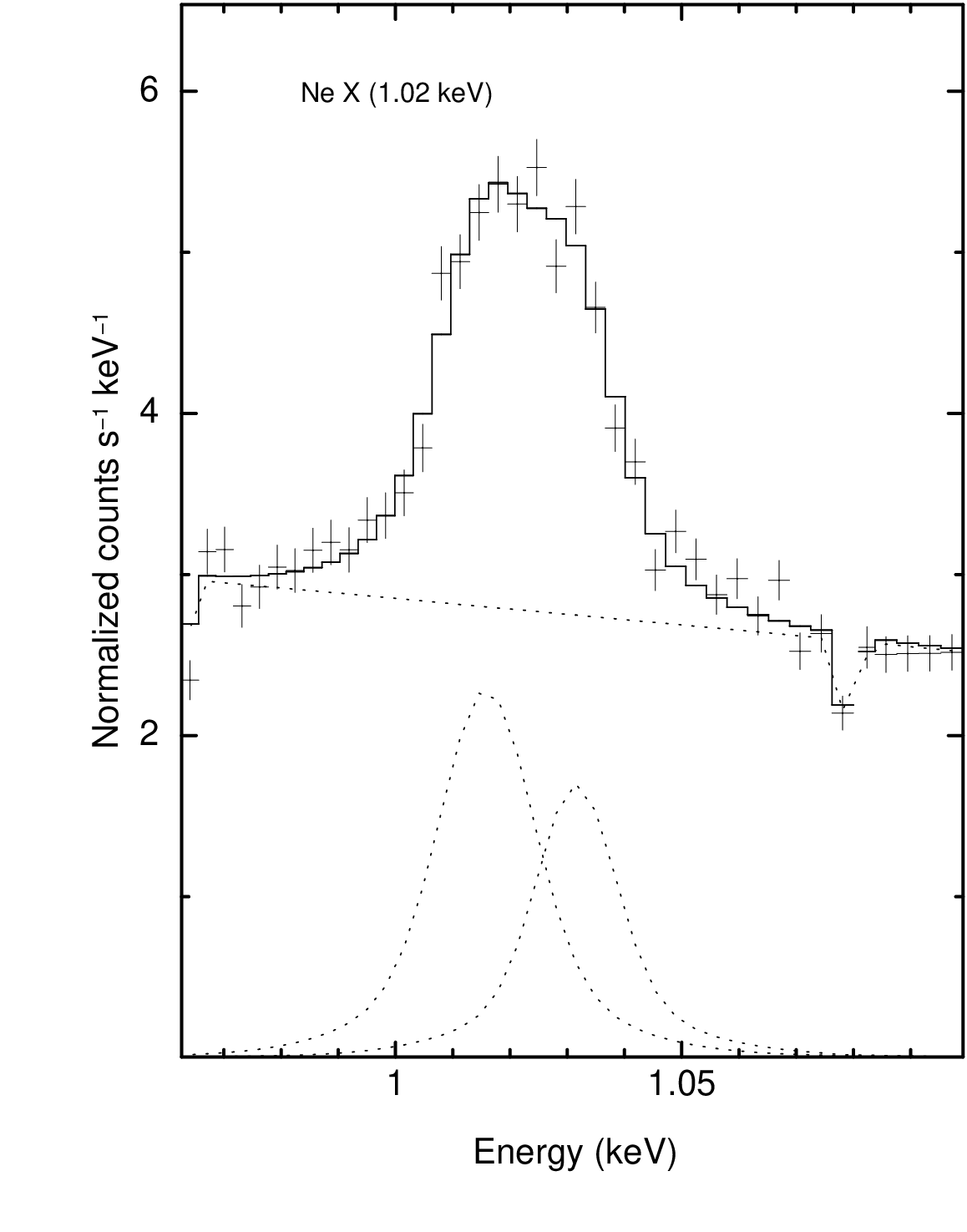}
\end{minipage}
\caption{Double-peaked emission lines.Left:~the hydrogenic O~VIII emission line observed with \textsc{RGS}-1 \& 2.
Right plot shows the broad hydrogenic Ne~X emission line as observed with the \textsc{RGS-2} data.}
\label{Double-peak}
\end{figure*}

 It is believed that Ne/O emission lines observed in the X-ray spectra of
 4U~1626--67 originate from highly ionized layers of the accretion disk.
The existence of double-peaked profiles support their disk
 origin \citep{Schulz01, Krauss07}.~Interestingly, we noticed that one of the emission lines 
 at 0.653~keV~(O~VIII) showed the presence of double-peaked profiles
 in the high resolution data of \textsc{RGS} (Figure~\ref{Double-peak}).
 Therefore, we fit this line with a pair of Gaussian
 to resolve into the doppler pairs and to estimate the disk velocities
 of red and blue-shifted components.
 The line velocities measured using the \textsc{RGS} data 
 along with the previous known values are given in Table~\ref{Velocity}.
The single-Gaussian fit revealed a broad emission line
 at 1.02~keV~(Ne~X) in the \textsc{RGS2} spectrum.~Therefore, we fit this line as well
 with a pair of Gaussian and the velocities measured are given in Table~\ref{Velocity}. 
 Ne~IX emission line observed in the \emph{XMM-Newton} data did not allow us
to measure doppler velocities. \\

\begin{figure*}
\centering
\begin{minipage}{0.45\textwidth}
\includegraphics[height=4.5in,width=7.5cm]{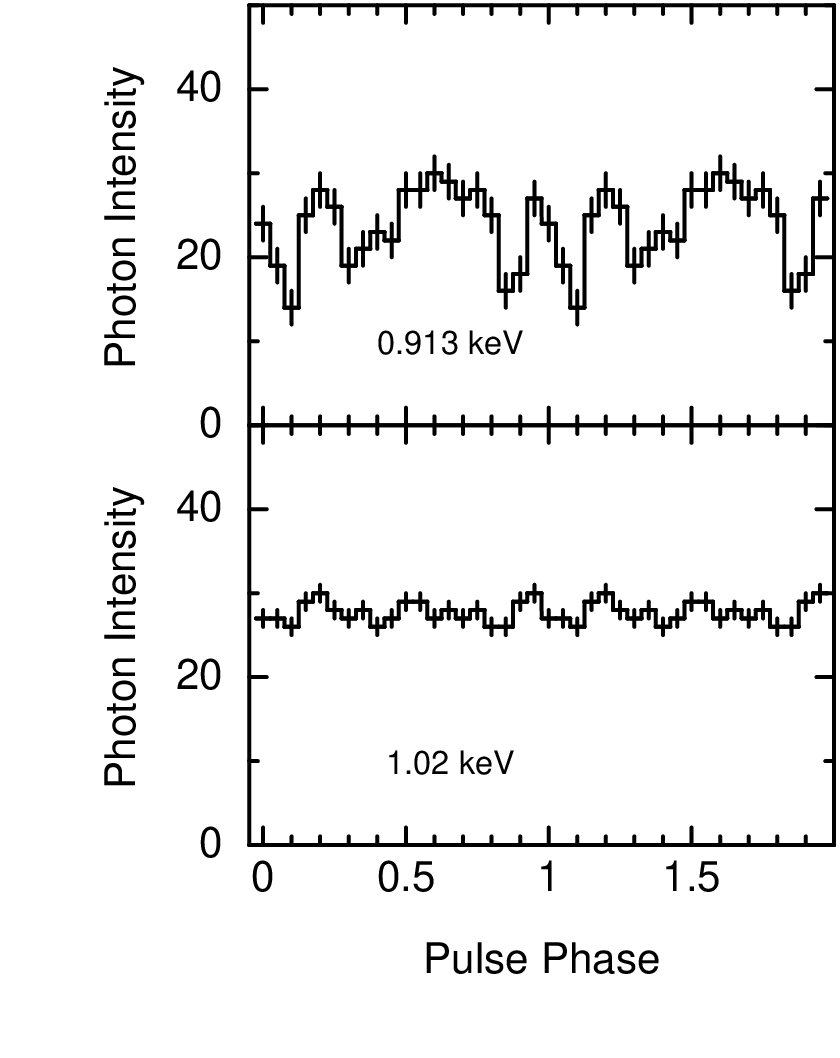}
\end{minipage}
\hspace{0.05\linewidth}
\begin{minipage}{0.45\textwidth}
\includegraphics[height=4.5in,width=7.5cm]{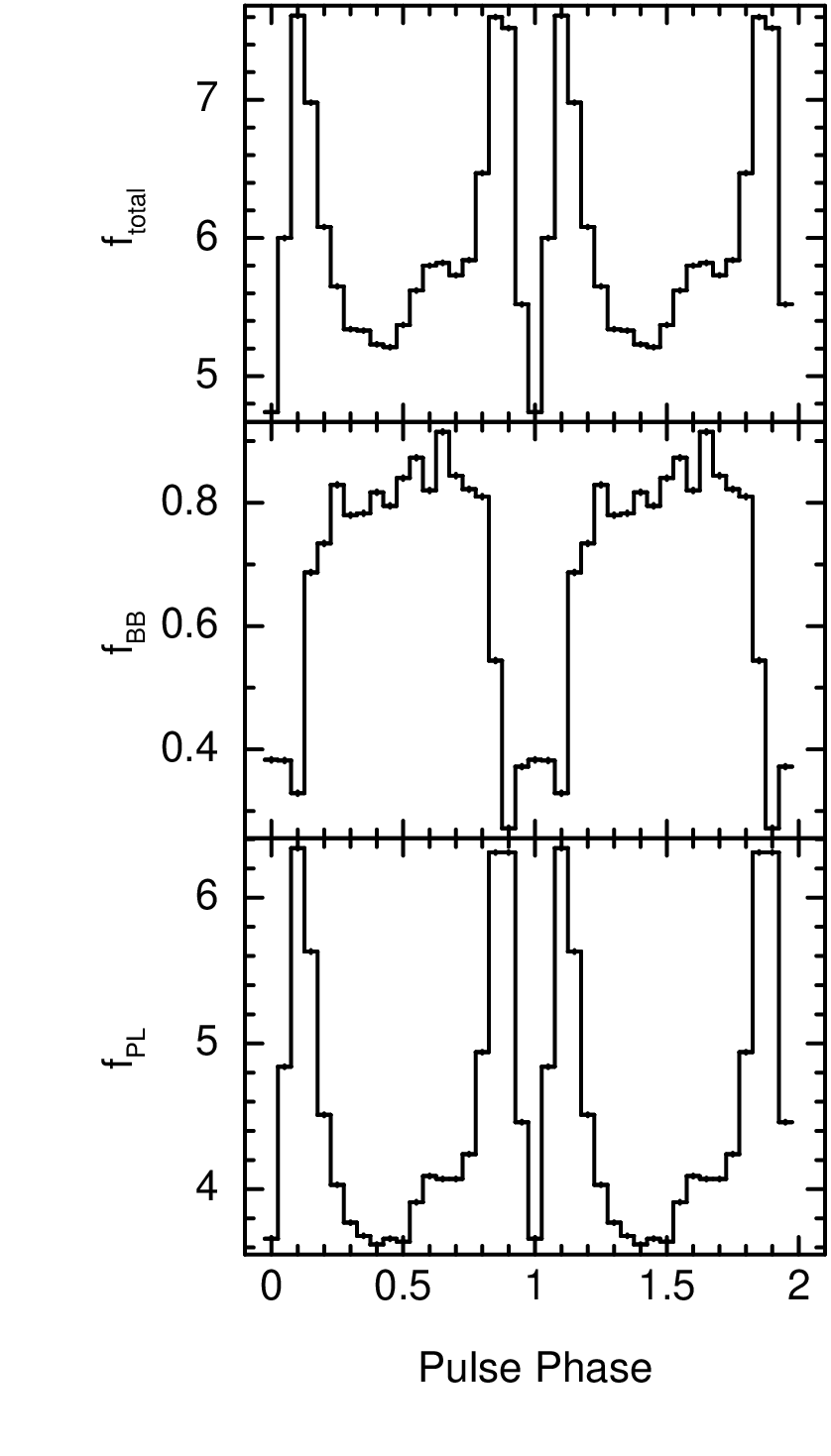}
\end{minipage}
\caption{Variation of line flux across the pulse phase is plotted in left 
while the plot in right shows the variation of continuum flux with pulse phase.
Line fluxes are in units of $10^{-4} photons/cm^{2}/sec$ while all the continuum fluxes~($f_{total}$,~$f_{BB}$,~$F_{PL}$)
are measured in $10^{-10} ergs/cm^{2}/sec$.
~All the errors are quoted with 1~$\sigma$ confidence.}
\label{pp-flare}
\end{figure*}

\begin{figure}
\centering
 \includegraphics[height=4.5in,width=7.5cm,keepaspectratio]{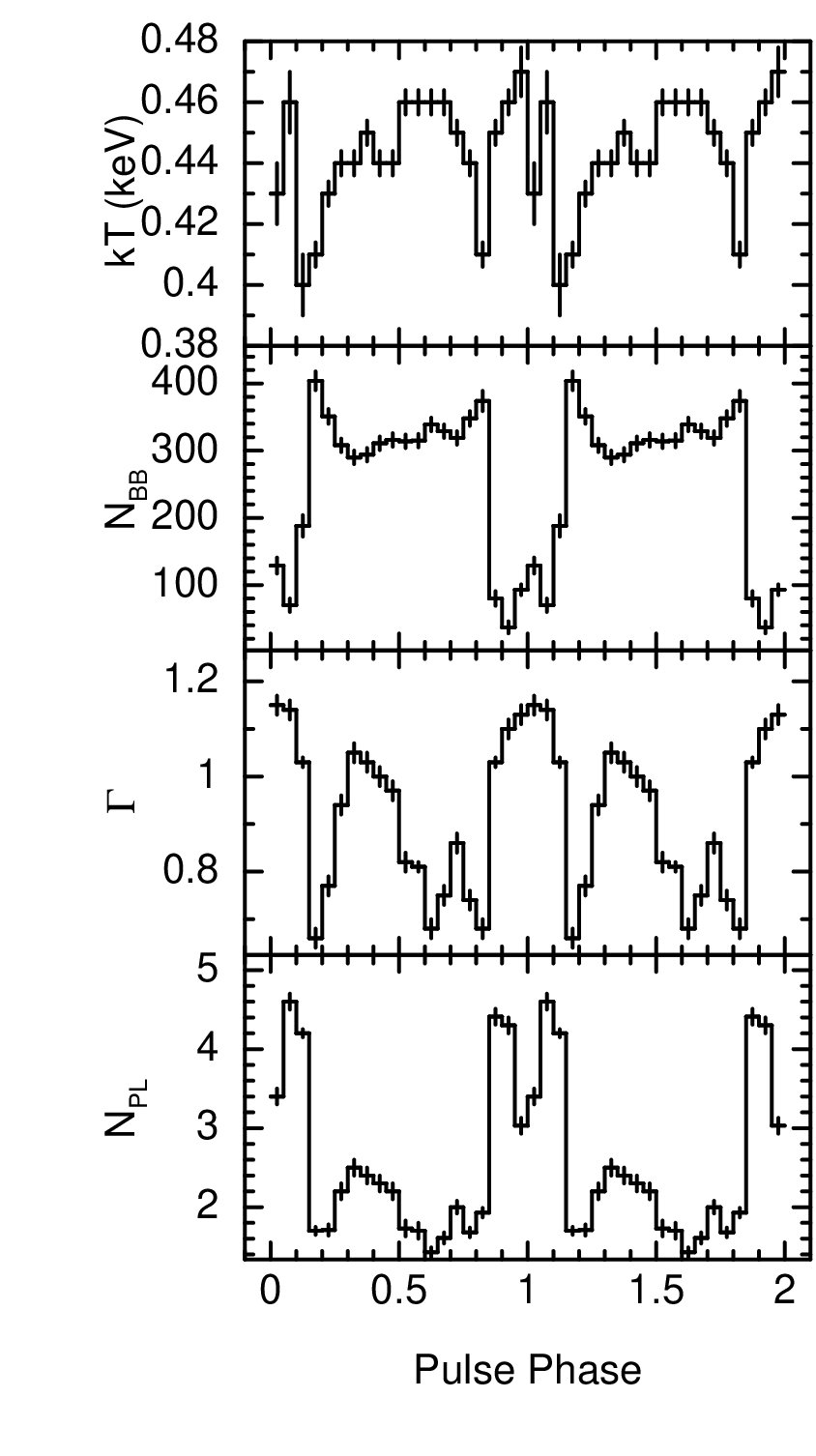}

\caption{
In this plot we show 
variation of continuum parameters across the pulse phase.}
\label{EW}
\end{figure}

  \subsection{Pulse Phase Resolved Spectroscopy}
  For performing pulse phase resolved spectroscopy 
  we have used data from \textsc{EPIC}-pn.
  We 
  added  $\textquoteleft$\textsc{phase}' column to the 
  pn event list.~This was performed using SAS task \textsc{phasecalc}
  with phase zero fixed at the reference time~(epoch) used for creating pulse profiles.
  Thereafter, appropriate good time intervals~(GTI) files
  were created for narrow phase bins of 0.05.
  These GTI files were used for the extraction of 20 phase resolved source spectra.
  Response matrices and ancillary response files used for
  phase averaged spectroscopy were used again for performing phase resolved spectroscopy.
  Spectral fitting was done in the energy range of 0.8-12~keV with the same spectral
  model consisting of a power-law, a blackbody and several emission lines. 
  We fixed the neutral hydrogen column density, line energies and line widths 
  to the values obtained from the phase averaged spectrum.
  Left plot of Figure~\ref{pp-flare} shows variation of flux of
low energy emission lines with pulse phase.
 From the plot we observe~:
\begin{itemize}
\item Ne~IX~$He_{\alpha}$ emission line at 0.913~keV shows strong variation
with pulse phase (factor of $\sim$ $2.0\pm0.3$).
~$\chi^2$ value of 103 for 20 phase bins
was observed after fitting a constant to the flux of line at 0.913~keV.
\item Ne~X~$Ly_{\alpha}$ emission line at 1.02~keV shows no significant variation with pulse phase.
This is similar to the previous observations made during spin-down phase of 4U~1626-67 \citep{Angelini95,Beri15}.
A constant fitted to the flux of Ne~X~$Ly_{\alpha}$ emission line showed a $\chi^2$ value of 32 for 20 phase bins. 

\end{itemize}

We estimated the observed total continuum flux in the energy band of 0.7-12~keV,
\textit{power law} flux in 2-12~keV band and the \textit{blackbody}
flux in 0.7-2~keV band using the \textsc{CFLUX} convolution model
in \textsc{XSPEC}.
The continuum flux profile plotted in right hand side of Figure~\ref{pp-flare}
shows that the modulation of the \textit{power law} is same as the modulation of the 
total flux while the flux modulation of the 
\textit{blackbody} component has a different shape.
The blackbody component shows a broad dip, 
consistent with the pulse profile in the 0.3-2.0~keV band 
in which the blackbody component dominates.~Shape 
of the power law profile can be imagined to have formed 
as a narrow dip at the centre of a broad pulse peak, while the
blackbody profile is a broad dip on an otherwise constant emission. \\

Continuum parameters also showed variation with pulse phase
(Figure~\ref{EW}).
Blackbody temperature shows strong variation 
with possible correlation with the pulse profile.
However, blackbody normalization profile shape
is anti-correlated to its temperature profile.
Power law index profile shows a sharp dip 
at phase 0.2 with some structures in rest of the profile
while power law normalization shows strong correlation
with the pulse profile shape~(bi-horned peaks around pulse phase 0.9 and 1.1).
The blackbody flux is a few percent of the total flux and given 
the systematic errors in \emph{EPIC}-\textsc{pn}, one should
be cautious about the blackbody parameters. The flux modulation
of the blackbody is however certainly different from the power law flux
variation, as is evident from the energy-resolved pulse profiles.
\begin{figure*}
\centering
\includegraphics[height=3.5in,width=7.5cm]{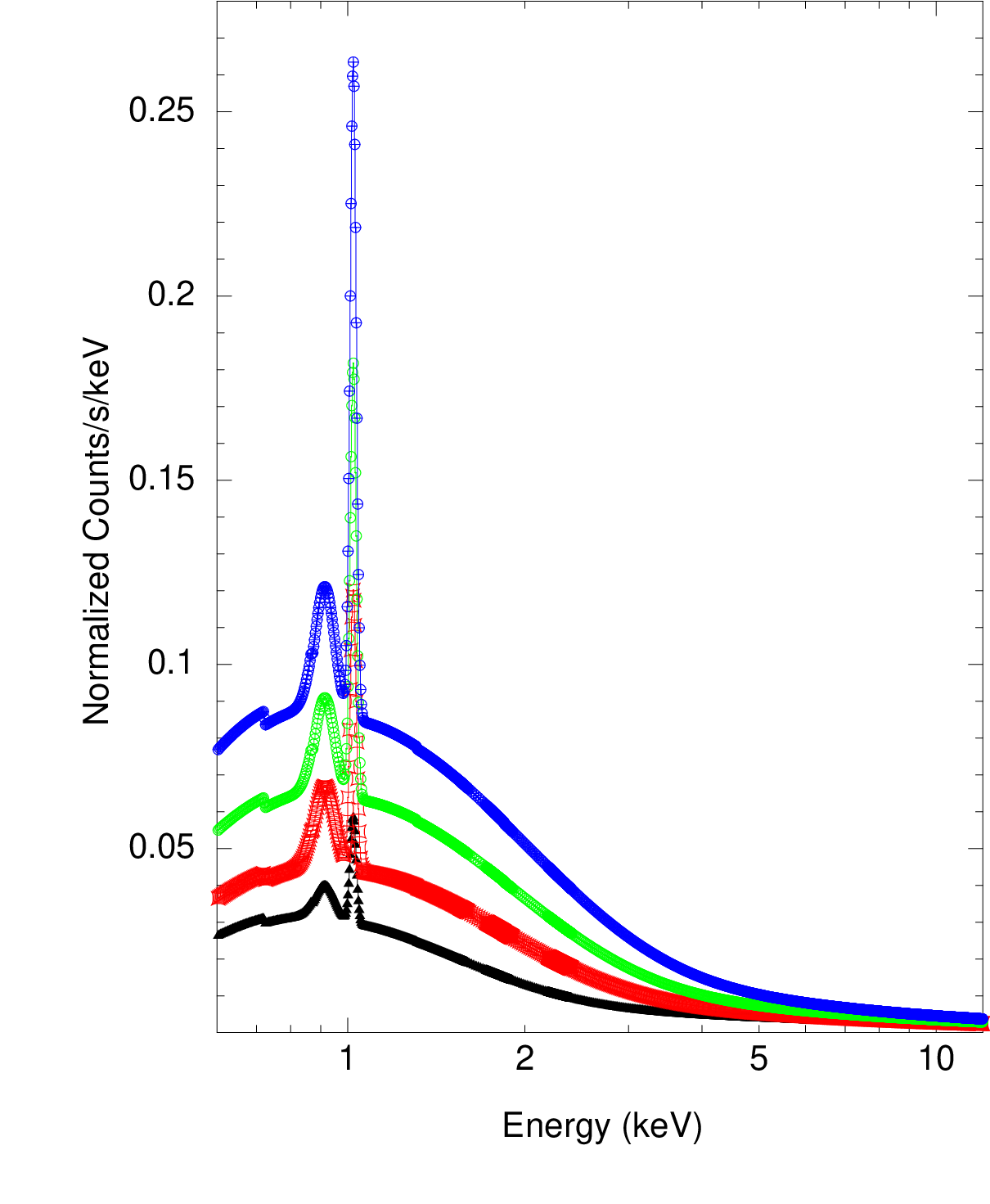}
\caption{Best fit model components for Intensity Resolved Spectra. Blue colour represents the spectrum created using intensity range of 200-320~counts/s,
green : 150-200~counts/s, red for 90-100~counts/s and black for spectrum during dip.~From the figure it is clear that normalisation varies
with the change in intensity.}
\label{Int-resol-spec-XMM}
\end{figure*}

\begin{figure*}
\centering
\begin{minipage}{0.45\textwidth}
\includegraphics[height=5.5in,width=7.5cm,keepaspectratio]{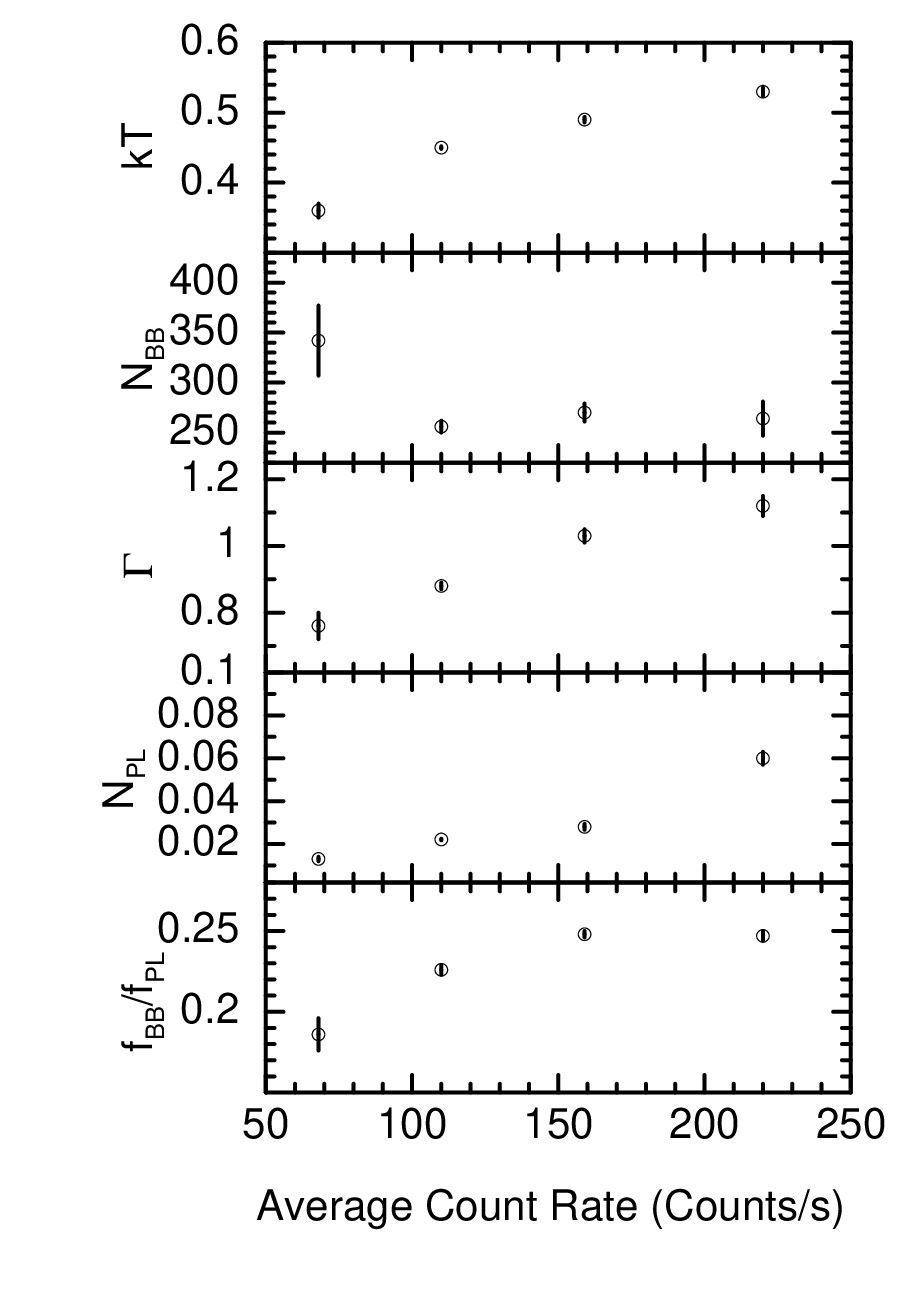}
\end{minipage}
\hspace{0.05\linewidth}
\begin{minipage}{0.45\textwidth}
\includegraphics[height=4.in,width=6.5cm,keepaspectratio]{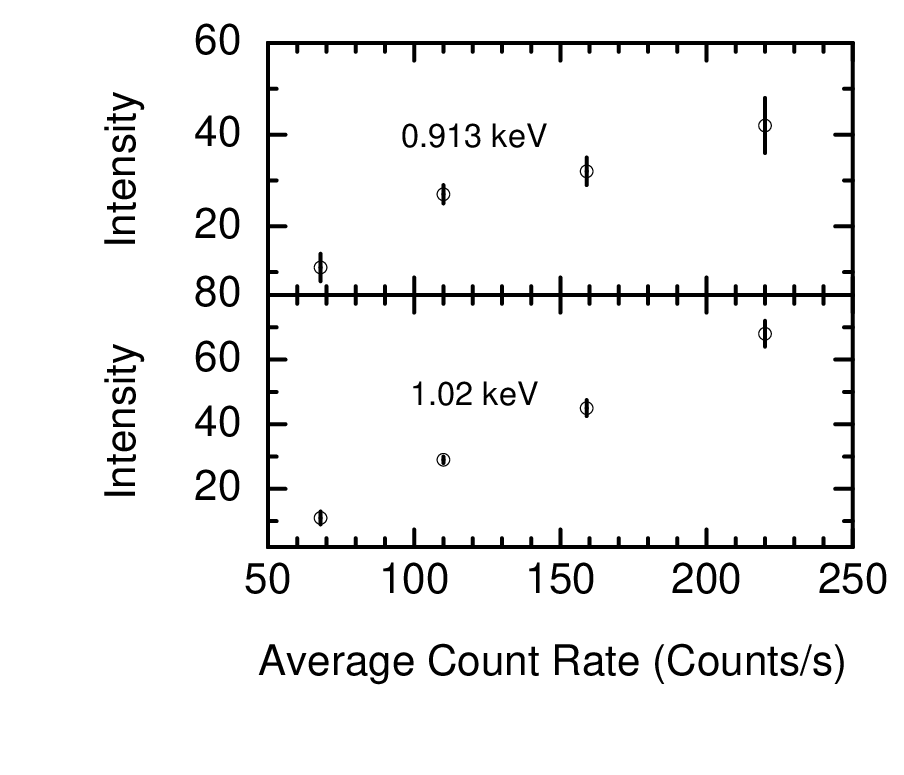}
\includegraphics[height=4.in,width=6.5cm,keepaspectratio]{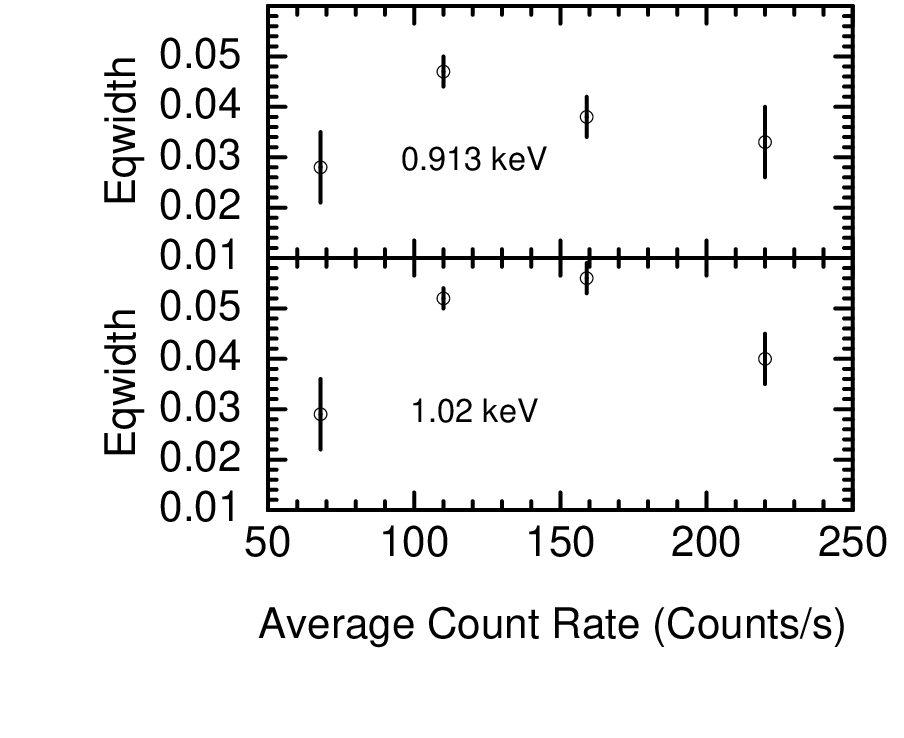}
\end{minipage}
\caption{Left hand plot shows the variation of continuum parameters with the increase in intensity, the bottom panel of this plot
shows the variation of ratio of blackbody flux and the power law flux.~On the right: Top plot shows the
variation of line flux with the increase in intensity.~Line fluxes are in units of $10^{-4} photons/cm^{2}/sec$
while the bottom plot shows the variation of equivalent width of Ne~IX and Ne~X emission lines 
with the increase in intensity.
~Note: X-axis label refer to average count-rate during different 
intensity resolved spectra.~All the errors are quoted with 1~$\sigma$ confidence.}
\label{Int-variation-flare}
\end{figure*}
\subsection{Intensity Resolved Spectroscopy}
Since strong variation in count-rates is observed
in light curve shown in Figure~\ref{LC},
we extracted intensity resolved 
spectra from \textsc{EPIC}-pn data using SAS task \textsc{evselect}.
Good~time~intervals~(GTIs) were created in different intensity ranges~:~sharp dip seen just after the end
of first flare seen in Figure~\ref{LC} and pn count-rates between 90-150~c/s,
 150-200~c/s,~200-320~c/s were used to extract intensity resolved spectra.
For spectral fitting we used the same technique which we opted while performing
phase resolved spectroscopy.~Neutral hydrogen column density,~line energies
and widths were fixed to the phase averaged values.~We also added $2\%$ systematics
while performing the spectral fitting.
Fitted model components are shown in Figure~\ref{Int-resol-spec-XMM}.
Temperature of blackbody and the power-law index was found to increase.
The line fluxes and fractional contribution of the blackbody flux
also found to increase with total flux (Figure~\ref{Int-variation-flare}).

\begin{table*}
\caption{Radius measurement for the Ne/O line formation region}
\label{table2}
\begin{tabular}{ c c c c c c c c c}
\hline
\hline
Ion Species & $Radius^{a}$~(Spin-down)  & $Radius^{a}$~(Spin-up) \\
 \hline
  O~VII & 2.3 & 4.2   \\
 O~VIII & 1.4 & 2.6  \\
 Ne~IX & 1.2 &  2.1 \\
 Ne~X &  0.07 & 0.14 \\
\hline
\end{tabular}
\label{radii}
\\{Note:~a~$\rightarrow$ Radius Measurements are in units of $10^{10}~cm$ }
\end{table*}

\section{Discussion \& Summary of Results}
In this paper we present results obtained using data from the
\emph{XMM-Newton} observatory during the current spin-up phase of
4U~1626--67.~Several new
and significant changes have been observed in comparison to the previous
observation made during its spin-down phase.~The main focus
of our study is to observe a pulse phase dependence of low energy
emission lines seen in the X-ray spectrum of 4U~1626--67.
Strong pulse phase dependence of O~VII emission line at 0.569~keV
was observed during spin-down phase of 4U~1626--67.~This strong variation was
interpreted as a result of warps in the accretion disk \citep{Beri15}.
Dissimilarities in timing characteristics (such as pulse profile, QPOs)
during spin-down and spin-up eras are believed to be due to difference in the
inner accretion flow from a warped accretion disk during spin-down phase \citep{Beri14,Kaur08}.
Therefore, one expects to see a different behavior of line fluxes with
pulse phase during current spin-up phase of this source. 
The calibration issues below 0.7~keV in the timing mode data of \textsc{EPIC}-pn 
did not allow us to study the pulse phase dependence of emission line features
at 0.571~keV~(O VII), 0.653~keV~(O~VIII) and 0.733~keV~(Fe~L).~However, we investigated the behavior
of emission lines at 0.913~keV~(Ne~IX) and 1.02~keV~(Ne~X) with the pulse phase. \\

We summarize the results as follows~:
\begin{itemize}
 \item Light curve obtained using the \textsc{EPIC}-pn data during its current spin-up phase
 showed dips.~Unlike other flaring sources
 like LMC~X-4, SMC~X--1 it is interesting to notice a broad dip in the light curve 
 soon after the decay of a large flare.~This feature is similar to that
 observed in the bursting pulsar, GRO~J1744-28.
 The light curve of GRO~J1744-28 showed the presence of
 a dip and recovery period following each outburst \citep[see e.g.,][]{Giles96}
 and the X-ray spectrum of GRO~J1744-28 showed no significant change going from quiescence to outburst \citep{Cannizzo96}.
The same authors proposed that the outbursts observed in the bursting pulsar
 could be due to Lightman-Eardley~(LE)~instability \citep{Lightman74} in the accretion disk
 and the material that is evacuated
onto the pulsar during an accretion event is replenished by
 material flowing in from further out; hence, the dip and recovery in the light curve following an outburst.
 After performing intensity resolved spectroscopy of 4U~1626--67 we found that overall there is no change in the shape of
 shape of the spectrum (see Figure~\ref{Int-resol-spec-XMM}) except that the continuum and line parameters
 follow an increasing trend with intensity.
Therefore, it is plausible that a similar mechanism might be responsible for the presence of flares, sharp dips 
in the light curve of 4U~1626--67.  \\

 
 \item QPO feature around 3~mHz is observed in the PDS. This feature has been observed for the first time using X-ray data
 of current spin-up phase.~The feature at 3~mHz also shows strong energy dependence.
 The feature is sharp at low energies as flares dominate at low energies.

  The energy dependence of fractional rms amplitude of QPO has been used as a tool
 to understand the physical origin of QPO \citep[see e.g.,][]{Gilfanov03,Cabanac10,Mukherjee12}.
The fractional rms amplitude of 3~mHz QPO
observed in 4U~1626--67 showed an increase
 upto 5~keV and thereafter its value saturates~(probably due to lower count rates at higher energies).
 Therefore, it seems that fluctuations in the blackbody component could be a plausible cause of the
 observed mHz QPO in 4U~1626--67.
 
 \item We found that the pulse profile shape below 2~keV
 is different from that seen during spin-down phase of 4U~1626--67 (Figure~\ref{PP-XMM}).
 Moreover,~during its spin-up phase pulse profiles below 2~keV 
are quite different from that seen above 2~keV (Figure~\ref{EN-pp}).
A possible explanation to these observations is 
changes in the emission diagram of the accretion column. 
During the low luminosity phase~(spin-down) of 4U~1626--67, the emission
of the accretion column is concentrated in a beam, oriented along the magnetic 
field axis while during the high luminosity phase~(spin-up) the emission diagram
changed to the fan beam pattern \citep{Basko75}.~Soft X-ray emission~(below~2~keV) is 
attributed to reprocessing of the primary emission by the optically thick material (i.e., the inner accretion disk)
and, therefore, changes in the emission diagram might lead to the changes in the
illumination of the inner accretion disk and hence, different pulse profiles below 2~keV
during the spin-up phase compared to the spin-down phase of 4U~1626---67.
We also note that the similar hypothesis was also proposed by \citet{Koliopanos16} to explain
the origin of the iron line during spin-up phase.

 
 \item Values of EW of emission lines observed in the phase averaged spectrum
 suggests that the EW of O~VIII has increased by a factor of 4
compared to the value~($\sim$7~eV) measured with \emph{Suzaku} during its spin-up phase
by \citet{Camero12}.~However, EW measured with \emph{ASCA} and \emph{XMM-Newton}
during its spin-down phase was $\sim$~14~eV \citep{Angelini95,Krauss07}.
Observations made with \emph{ASCA} and \emph{XMM-Newton} during spin-down phase of 4U~1626-67 revealed 
EW of O~VII to be $\sim$31~eV and $\sim$23~eV respectively \citep{Angelini95,Krauss07}
while the measurement made during spin-up phase with the \emph{Suzaku} observatory showed
a much lower value (1.3~eV).
EWs of Ne~IX and Ne~X emission lines are almost consistent with the previous measurements
made during its spin-up phase \citep{Camero12}. \\

\item From the intensity resolved spectroscopy, we found that there is an increase in the ratio
between blackbody and power law flux which suggests that the spectrum 
softens with the intensity.~The values of line fluxes at 0.913~keV and 1.02~keV also showed an
increase with intensity.
However, we did not notice any correlation between the equivalent width of these emission lines
with the intensity
(Figure~\ref{Int-variation-flare}). \\

 \item 
 
From the pulse phase resolved spectroscopy of 4U~1626--67, we observed a strong variation of
Ne~IX emission line with the pulse phase while the emission line at 1.02~keV~(Ne~X)
showed a lack of pulsations.
A different behaviour of Ne~IX and Ne~X emission lines accross the pulse phase
suggests that these emission lines might have a different origin.~It may be possible that
Ne~IX emission line originates from the accretion disk and thus, showing a strong pulse phase
dependence while the Ne~X emission line originates from highly ionized optically thin
emission i.e. from the material trapped in the Alfven shell \citep{Basko80}.
If this scenario is true, it also provides an explanation to the different line shapes 
of Ne~IX and Ne~X emission lines.
The broadening observed in the profile of Ne~IX emission line could be due to the Doppler shifts while
the microscopic processes may be the cause of the broadening of Ne~X emission line.  \\

Current observation made during spin-up phase of 4U~1626--67 showed a
different line intensity modulation pattern of Ne~IX emission line compared to the earlier \emph{XMM-Newton} 
observation in the spin-down phase.
Pulse phase dependence of low energy emission lines in 4U~1626--67 is believed to be due to
geometrical effect called ``warping{\textquotedblright} of the accretion disk \citep[][]{Beri15}.
Due to warps~(wherein tilt angle of the normal to the local disk surface varies with azimuth) 
in the accretion disk, one expects to observe modulation in the flux
of reprocessed emission visible along our line of sight.
Several possibilities have been discussed in the literature that might lead to
warps in the accretion disk.
One of the widely accepted possibility is that if the accretion disk is subject
to strong central irradiation, then it is unstable to warping \citep[see e.g.,][]{Petterson77,Pringle96}.
We also note that \citet{Pringle96} suggested that radiation-driven warping is
strongest in the outer regions of the accretion disks.
 From Table~\ref{Velocity}, it is interesting to notice that O~VIII emission line at 0.653~keV 
 showed a lower value of velocity compared to the values measured during the
 spin-down phase of 4U~1626--67.
 This indicates that the radius of the accretion disk at which this emission line is formed
 has moved outward during spin-up phase. \\
 
 In order to further investigate the above,
  we estimated radii of Ne/O emission line formation regions
 using the expression for ionization parameter~($\zeta$~=$~L_X/nR^2$), where $L_X$ is the 
 X-ray luminosity, $n$ is the ion number density and $R$ is the radius (see Table~\ref{radii})
 The values of ionization parameters calculated using the XSTAR code~\citep{Kallman82} for the optically thin
 photoionized model \footnote{http://heasarc.gsfc.nasa.gov/lheasoft/xstar/xstar.html}
 were opted for our calculations.
 We further assumed a constant value~($10^{13}~cm^{-3}$) for the electron number density.
 This is a reasonable assumption as comparable values of number density were estimated 
 by \citet{Schulz01}.
 It is interesting to notice from Table~\ref{radii} that the radius of line formation region for each of the ion
 species has moved outwards compared to the values obtained using $L_X$ measured during spin-down phase.
 This further supports our interpretation that 
 a strong variation of Ne~IX line during 
the current spin-up phase of 4U~1626--67 is because the structures~(or warps) in the accretion disk (that produce pulse phase dependence of emission lines) 
have changed during its spin-up phase or the line forming region has moved outwards
where the warps dominates. \\
 
Different pulse phase dependence of Ne~IX emission line observed during current spin-up phase of 4U~1626--67,
therefore, supports that there is a possible change in accretion flow geometry.
Accretion flow geometry plays an important role in transfer of angular momentum and therefore
any change in it would suggest a change in the interaction between the Keplerian
disk and the stellar magnetic field at the corotation radius. 
\end{itemize}

\section*{Acknowledgments}
We thank the anonymous referee for several useful suggestions which improved the
quality of the paper.
A.B. gratefully acknowledge Raman Research Institute (RRI) for providing local hospitality
and financial assistance, where this work was started.
She is also grateful to the Royal Society and SERB~(Science $\&$ Engineering Research Board, India)
for financial support through Newton-Bhabha Fund.
A.B would like to extend further thanks to 
Michael~Smith and Matteo~Guainazzi for their useful
insights about \emph{XMM-Newton} data analysis.
The authors would like to thank all the members of the
\emph{XMM-Newton} observatory for carrying out observation
of 4U~1626--67 during its current spin-up phase and 
for their contributions in the instrument preparation,~spacecraft operation,~software development,~and~ 
in-orbit instrumental calibration.
\bibliography{1626}{}
 \bibliographystyle{mnras}
 
\label{lastpage}
\end{document}